\documentclass[12pt,psfig,epsf]{article}
\usepackage{psfig}
\renewcommand{\theequation}
{\arabic{section}.\arabic{equation}}

\makeatletter
\def\eqnarray{ \stepcounter{equation} \let\@currentlabel=\theequation
 \global\@eqnswtrue
 \global\@eqcnt\z@
 \tabskip\@centering
 \let\\=\@eqncr
 $$\halign to \displaywidth\bgroup\@eqnsel\hskip\@centering
 $\displaystyle\tabskip\z@{##}$&\global\@eqcnt\@ne
 \hfil$\displaystyle{{}##{}}$\hfil
 &\global\@eqcnt\tw@$\displaystyle\tabskip\z@{##}$\hfil
 \tabskip\@centering&\llap{##}\tabskip\z@\cr}
\makeatother

\makeatletter
\def\@arrayacol{\edef\@preamble{\@preamble \hskip .5\arraycolsep}}
\def\array{\let\@acol\@arrayacol \let\@classz\@arrayclassz
\let\@classiv\@arrayclassiv \let\\\@arraycr\def\@halignto{}\@tabarray}
\makeatother
\makeatletter
\newcounter{subeqncnt}
\def\thesubeqncnt{\alph{subeqncnt}}
\def\subequations{\begingroup%
   \stepcounter{equation}\edef\@tempa{\theequation}%
   \let\c@equation\c@subeqncnt\c@subeqncnt\z@
   \edef\theequation{\@tempa\noexpand\thesubeqncnt}}

\makeatother
\newcommand{\be}{\begin{equation}}
\newcommand{\ee}{\end{equation}}

\newcommand{\beqa}{\begin{eqnarray}}
\newcommand{\eeqa}{\end{eqnarray}}
\newcommand{\nn}{\nonumber}

\newcommand{\eqref}[1]{(\ref{#1})}

\def\CF {{\cal F}}

\def\CL {{\cal L}}

\def\CV {{\cal V}}

\begin{document}

\setlength{\baselineskip}{7mm}
\begin{titlepage}
\begin{flushright}

{\tt NRCPS-HE-13-09} \\
{\tt CERN-PH-TH/2009-126}\\

July, 2009
\end{flushright}

\vspace{1cm}

\begin{center}
{\it \Large Scattering of Charged Tensor Bosons    \\ in  \\
 Gauge   and Superstring  Theories
}

\vspace{1cm}

%\author{
{\sl Ignatios Antoniadis$^{1}$${} $\footnote{${}^\dagger$ On leave of absence
from CPHT {\'E}cole Polytechnique, F-91128, Palaiseau Cedex,France.} and  George Savvidy$^{2}$

\bigskip
\centerline{${}^1$ \sl Department of Physics, CERN Theory Division CH-1211 Geneva 23, Switzerland}
\bigskip
\centerline{${}^2$ \sl Demokritos National Research Center, Ag. Paraskevi, GR-15310, Athens}
\bigskip

}%author ends
%}
%\date{}%in order NOT to write the date
%\maketitle

\end{center}

\vspace{1cm}

\begin{abstract}

\end{abstract}
We calculate the leading-order scattering amplitude of one vector  and
two tensor gauge bosons in a recently proposed non-Abelian tensor gauge field theory and open superstring theory.
The linear in momenta part of the superstring amplitude has identical Lorentz structure
with the gauge theory, while its cubic in momenta part
can be identified with an effective Lagrangian which is
constructed using generalized non-Abelian field strength tensors.

\end{titlepage}

\pagestyle{plain}
%\pagenumbering{roman}

\section{\it Introduction}

An infinite tower of particles of high spin naturally
appears in the spectrum of different string theories. In the zero slope
limit massless states of open and closed strings
can be identified with vector - Yang-Mills and tensor - graviton gauge quanta
\cite{Neveu:1971mu,Yoneya:1974jg,Yoneya:2009ju,Scherk:1974mc,Scherk:1974ca,
Green:1987sp,Polchinski:1998rq}. Massive string states can be described by string
field theory developed in
\cite{Witten:1985cc,Thorn:1985fa,Siegel:1985tw,Siegel:1988yz,Arefeva:1989cp,
Taylor:2003gn,Taylor:2006ye}. Nevertheless to represent the Lagrangian and equations
in terms of components of tensor fields still remains a challenge \cite{Siegel:1985tw}.
In this respect higher spin field theories have
received large attention   \cite{schwinger,Weinberg:1964cn,
Weinberg:1964ev,Weinberg:1964ew,
chang,singh,singh1,fronsdal,fronsdal1,Bengtsson:1983pd,
Bengtsson:1983pg,Bengtsson:1986kh,Berends:1984rq,Curtright:1987zc}
together with the recent development of interacting field theories of
high spins
\cite{Bekaert:2005vh, Vasiliev:2005zu, Francia:2006hp,
Bouatta:2004kk,Sagnotti:2005ns,Bekaert:2006us,Bekaert:2005jf,Metsaev:2005ar}.

As we mentioned the massless states of {\it open}
superstring theory with Chan-Paton charges
\cite{Paton:1969je} have been identified with the Yang-Mills gauge quanta
\cite{Neveu:1971mu,Green:1987sp,Polchinski:1998rq,Kawai:1985xq,
Antoniadis:2007ta,Antoniadis:2009nv,Antoniadis:2006mr}
and it is of great importance to identify also   massive higher spin string states
with states of some Lagrangian quantum field theory.

One can imagine that massive states of open string may be described
by some extension of Yang-Mills theory  to non-Abelian
tensor gauge field theory. Such extension of Yang-Mills theory
which includes {\it charged tensor gauge fields} was suggested recently in
\cite{Savvidy:2005fi,Savvidy:2005zm,Savvidy:2005ki,Savvidy:2005vm}.
Not much is known about physical properties of this gauge
field theory and our intension  is to compare tree-level scattering amplitudes
of tensor gauge bosons in non-Abelian tensor gauge field theory and
open superstring theory.

Non-Abelian tensor gauge fields are defined as
rank-(s+1) tensor potentials
$
A^{a}_{\mu\lambda_1 ... \lambda_{s}}.
$\footnote{Tensor gauge fields
$A^{a}_{\mu\lambda_1 ... \lambda_{s}}(x),~~s=0,1,2,...$
are totally symmetric with respect to the
indices $  \lambda_1 ... \lambda_{s} $, but with no {\it a priori}
symmetry with respect to the first index  $\mu$.
In particular, we have
$A^{a}_{\mu\lambda}\neq A^{a}_{\lambda\mu}$ and
$A^{a}_{\mu\lambda\rho}=A^{a}_{\mu\rho\lambda} \neq A^{a}_{\lambda\mu\rho}$.
The adjoint group index $a=1,...,N^2 -1$
in the case of $SU(N)$ gauge group.}
The gauge invariant
Lagrangian describing dynamical tensor gauge bosons of all ranks
has the form \cite{Savvidy:2005fi,Savvidy:2005zm,Savvidy:2005ki,Savvidy:2005vm}
\be\label{Lagrangian}
{\cal L} ~= ~
{{\cal L}}_{YM} + {\cal L}_2 +  {\cal L}_3  +....,
\ee
where ${{\cal L}}_{YM}$ is the Yang-Mills Lagrangian and
defines {\it cubic and quartic interactions} with
{\it dimensionless coupling constant}.\footnote{In D-dimensions
the coupling constant has dimension $(4-D)/2$.}
For the lower-rank tensors, the Lagrangian has the following form
\cite{Savvidy:2005fi,Savvidy:2005zm,Savvidy:2005ki}:
\beqa\label{totalactiontwo}
{{\cal L}}_2  =&-&
{1\over 4}G^{a}_{\mu\nu,\lambda}G^{a}_{\mu\nu,\lambda}
-{1\over 4}G^{a}_{\mu\nu}G^{a}_{\mu\nu,\lambda\lambda}  \\
&+&{1\over 4}G^{a}_{\mu\nu,\lambda}G^{a}_{\mu\lambda,\nu}
+{1\over 4}G^{a}_{\mu\nu,\nu}G^{a}_{\mu\lambda,\lambda}
+{1\over 2}G^{a}_{\mu\nu}G^{a}_{\mu\lambda,\nu\lambda},\nn
\eeqa
where the generalized gauge field strengths are:
\beqa\label{tensors}
G^{a}_{\mu\nu} &=&
\partial_{\mu} A^{a}_{\nu} - \partial_{\nu} A^{a}_{\mu} +
g f^{abc}~A^{b}_{\mu}~A^{c}_{\nu}, \nn\\
G^{a}_{\mu\nu,\lambda} &=&
\partial_{\mu} A^{a}_{\nu\lambda} - \partial_{\nu} A^{a}_{\mu\lambda} +
g f^{abc}(~A^{b}_{\mu}~A^{c}_{\nu\lambda} + A^{b}_{\mu\lambda}~A^{c}_{\nu} ~), \\
G^{a}_{\mu\nu,\lambda\rho} &=&
\partial_{\mu} A^{a}_{\nu\lambda\rho} - \partial_{\nu} A^{a}_{\mu\lambda\rho} +
g f^{abc}(~A^{b}_{\mu}~A^{c}_{\nu\lambda\rho} +
 A^{b}_{\mu\lambda}~A^{c}_{\nu\rho}+A^{b}_{\mu\rho}~A^{c}_{\nu\lambda}
 + A^{b}_{\mu\lambda\rho}~A^{c}_{\nu} ~) .\nn
\eeqa
The Lagrangian forms $ {{\cal L}}_{s}$
for higher-rank  fields can be found in %the previous publications
Refs.\cite{Savvidy:2005fi,Savvidy:2005zm,Savvidy:2005ki}.

Here we shall focus our attention on the lower-rank
tensor gauge field $A^{a}_{\mu\lambda}$, which
decomposes into a symmetric tensor $T_S$ of spin two and an antisymmetric
tensor $T_A$, Poincar\'e dual of spin zero, charged gauge bosons \cite{Savvidy:2005ki}.
The Feynman rules for these propagating modes and their interaction vertices
can be extracted from the Lagrangian (\ref{totalactiontwo})
and allow  to calculate tree-level scattering amplitudes for processes involving
vector and tensor gauge bosons \cite{Savvidy:2005ki}.

In the spectrum of open superstring theory
with Chan-Paton charges there is also a massless vector gauge
boson $V$ on the first excited level and a rank-two massive tensor boson  $T_S$
at the second level.
The emission vertices for these states are defined as follows
(in the zero and -1 ghost picture for $V$ and $T_S$, respectively)
\cite{Green:1987sp,Polchinski:1998rq,Kawai:1985xq}:
\beqa
V:~~~& e_{\mu}(k) (\dot{X}^{\mu}
-2i \alpha^{'} k\cdot \psi \psi^{\mu} ) e^{ik  X} ~~~~~~~~~~~~~  &\alpha^{'} k^2 = 0\nn\\
T_S:~~~&\varepsilon_{\mu\nu}(k) \psi^{\{\mu}(\dot{X}^{\nu}
-2i \alpha^{'} k\cdot \psi \psi^{\nu\}} ) e^{ik  X} ~~~~~~~~~~~~~
&\alpha^{'} k^2 = -1 .
\eeqa
and allow to calculate different tree level scattering amplitudes
involving vector and tensor bosons $V$ and $T_S$.
{\it Our intension  is to compare tree-level scattering amplitudes
of tensor gauge bosons in the above non-Abelian tensor gauge field theory and in
open superstring theory.}

We have found that the {\it linear in momenta} part of the 3-point superstring
amplitude has similar Lorentz structure with the one in gauge theory defined by the Lagrangian
${{\cal L}}_2$ (\ref{totalactiontwo}) and that
the {\it cubic in momenta} part of the
superstring amplitude can be identified with an effective Lagrangian  $\CL_{\partial}$ (\ref{effectiveaction1})
and $ \CL^{'}_{\partial}$ (\ref{effectiveaction2}) constructed using non-Abelian field strength tensors (\ref{tensors}).
This result suggests that most probably non-Abelian tensor gauge field theory
describes a sub-sector of excited states of open superstring theory with higher helicities,
similar to a  Yang-Mills theory describing the first
excited state. More complicated amplitudes
should be analyzed in order to solidify this proposal.

\section{\it VTT Amplitude in Tensor Gauge Theory}
In {\it massless} tensor gauge field theory, the on-shell tensor-vector-tensor  amplitude VTT
is \cite{Savvidy:2008ks}
\be\label{tensorgaugeamplitude}
M_{gauge~theory}= \varepsilon_{\alpha\acute{\alpha}}(k_1) e_{\beta}(k_3)
\varepsilon_{\gamma\gamma^{'}}(k_2)~
\CF^{\alpha\acute{\alpha}\beta\gamma\acute{\gamma}}(k_1,k_3,k_2)~
\delta(k_1 + k_2 +k_3)
\ee
where
\beqa\label{intymsymmetricvertex}
\CF^{\alpha\acute{\alpha}\beta\gamma\acute{\gamma}}(k_1,k_3,k_2) =
 &+&k^{\beta}_{12} ~ (\eta^{\alpha\gamma}\eta^{\alpha^{'}\gamma^{'}} +
\eta^{\alpha\gamma^{'}} \eta^{\alpha^{'}\gamma} )  \\
&+&{1\over 4}k^{\alpha}_{23}~ (\eta^{\beta\gamma}\eta^{\alpha^{'}\gamma^{'}} +
\eta^{\beta\gamma^{'}} \eta^{\alpha^{'}\gamma} )  \nn\\
&+&{1\over 4} k^{\alpha^{'}}_{23}~ (\eta^{\beta\gamma}\eta^{\alpha \gamma^{'}} +
\eta^{\beta\gamma^{'}} \eta^{\alpha \gamma} ) \nn\\
&+&{1\over 4}k^{\gamma}_{31}~ (\eta^{\alpha\beta}\eta^{\alpha^{'} \gamma^{'}} +
\eta^{\alpha^{'}\beta} \eta^{\alpha \gamma^{'}} ) \nn\\
&+&{1\over 4}k^{\gamma^{'}}_{31}~ (\eta^{\alpha\beta}\eta^{\alpha^{'} \gamma } +
\eta^{\alpha^{'}\beta} \eta^{\alpha \gamma} )  ,\nn
\eeqa
where $k^2_i=0~~ i=1,2,3$  and  $k_{ij}= k_i -k_j$. It is important
that in this massless theory the momentum conservation
$\delta(k_1 + k_2 +k_3)$ has a nontrivial solution
$$
k_1  = (\omega, 0, 0, r),~~k_3  =
(\omega, 0, 0, r),~~k_2 = (-2\omega, 0, 0, -2r)
$$
$(\omega^2 = r^2)$ that can be deformed by a complex parameter $z$
\cite{Berends:1981rb, Dixon:1996wi,Parke:1986gb,
Witten:2003nn,Britto:2004ap,Britto:2005fq,Benincasa:2007xk},
\be\label{deformation1}
k_1  = (\omega, z, iz, r),~~k_3  =
(\omega, -z, -iz, r),~~k_2 = (-2\omega, 0, 0, -2r).
\ee
Thus, the above expression for VTT
$\CF^{\alpha\acute{\alpha}\beta\gamma\acute{\gamma}}(k_1,k_3,k_2)$
 has a {\it nonzero phase space of validity} which is
parameterized by the complex parameter $z$ (\ref{deformation1}).
Depending on polarizations of scattered particles one can see that
there are only four non-zero helicity amplitudes
$M(+2,+1,-2)$,~  $M(-2,+1,+2),$ ~
$M(+2,-1,-2)$ and $M(-2,-1,+2)$.

Contracting Lorentz indices one can see that in {\it tensor gauge theory} the
amplitude (\ref{tensorgaugeamplitude}), (\ref{intymsymmetricvertex}) can explicitly be written in the form
\beqa\label{ymsymmetricvertexa}
&M_{gauge~theory}=\varepsilon_{\alpha\acute{\alpha}}(k) e_{\beta}(p)  \varepsilon_{\gamma\gamma^{'}}(q)
\CF^{\alpha\acute{\alpha}\beta\gamma\acute{\gamma}}(k,p,q) \delta(k + p + q)= \nn\\  &  = ~
4(k  \cdot e_p)  ~~(\varepsilon_k  \cdot \varepsilon_q)
 - 2(p  \cdot \varepsilon_k \cdot \varepsilon_q \cdot e_p)
 +   2 (e_p  \cdot \varepsilon_k \cdot \varepsilon_q \cdot p).
\eeqa
Furthermore, one should take into account that
all particles are massless $k^2=p^2=q^2 =0$ and that the momentum
conservation $k+p+q=0$ gives
$
k  \cdot p =p \cdot q=q \cdot k=0.
$
Thus the VTT amplitude is nontrivial only if one considers
complex momenta (\ref{deformation1}) or the space-time signature $\eta^{\mu\nu}=(-+-+).$
The polarization vectors and tensors we shall take are then in the form
$$
e^+_{k}={1\over \sqrt{2}} ({z \over \omega}, 1, -i, -{z \over r} ),~~
e^+_p={1\over \sqrt{2}}(-{z \over \omega}, 1, -i, {z \over r} ),~~
e^-_q={1\over \sqrt{2}}(0, 1,  i ,0 )
$$
where $\varepsilon_{\alpha\acute{\alpha}}(k)= e_{\alpha}(k) e_{\acute{\alpha}}(k)$ and
$\varepsilon_{\gamma\acute{\gamma}}(q)= e_{\gamma}(q) e_{\acute{\gamma}}(q)$.
We can calculate now the amplitude  (\ref{ymsymmetricvertexa}) using the
relations $e_k \cdot e_q = e_p \cdot e_q=1$,~~ $k \cdot e_p=
-p \cdot e_k = 2\sqrt{2}z$ and $e_k \cdot e_p=0$
\beqa\label{ymantisymmetricvertexspinorial1}
&M(+2,+1,-2)_{gauge~theory}= \\
&4(k  \cdot e_p)  ~~(\varepsilon_k  \cdot \varepsilon_q)
 - 2(p  \cdot \varepsilon_k \cdot \varepsilon_q \cdot e_p)
 +   2 (e_p  \cdot \varepsilon_k \cdot \varepsilon_q \cdot p)
=   12 \sqrt{2} z ,\nn
\eeqa
so that {\it the VTT amplitude has non-trivial analytical continuation
and is proportional to the deformation parameter z}.
In the same way one can compute other polarization amplitudes.

\section{\it Mass-shell gauge invariance of VTT in tensor gauge theory}

The expression for the VTT  amplitude (\ref{ymsymmetricvertexa}) is on {\it mass-shell gauge invariant}
\beqa\label{variationa}
&e_{\beta}(p) \rightarrow e_{\beta}(p) +  \xi ~ p_{\beta}\nn\\
&\varepsilon_{\alpha \alpha^{'} }(k) \rightarrow \varepsilon_{\alpha  \alpha^{'} }(k)
+ k_{\alpha} \xi_{\alpha^{'}} + k_{\alpha^{'}} \xi_{\alpha},~~~~~~k^2=0,~~~~k \cdot \xi =0
\eeqa
where $\xi$ and $\xi_{\alpha}$ are gauge parameters.
Indeed the gauge variation  of (\ref{ymsymmetricvertexa}) under
(\ref{variationa}) $\delta e_p \sim p$ is
\be
  \delta M_{gauge~theory} =
  4 (k \cdot p) ~ (\varepsilon_k \cdot \varepsilon_q)
- 2(p  \cdot \varepsilon_k \cdot \varepsilon_q \cdot p)
 +   2 (p  \cdot \varepsilon_k \cdot \varepsilon_q \cdot p) =0
\ee
and under (\ref{variationa}) $\delta \varepsilon_k \sim k \otimes \xi + \xi \otimes k$ is
\beqa
& \delta M_{gauge~theory} = 8 (k \cdot e_p) ~  (k \cdot \varepsilon_q \cdot \xi)-\\
&- 2 (p  \cdot  k) ~ (\xi \cdot \varepsilon_q \cdot e_p)
- 2 (p  \cdot \xi) ~ (k  \cdot \varepsilon_q \cdot e_p)
 +  2  (e_p  \cdot k) ~(\xi \cdot \varepsilon_q \cdot p)
 +  2  (e_p  \cdot  \xi) ~(k \cdot \varepsilon_q \cdot p)=0,\nn
\eeqa
because $p  \cdot  k =0,~~  k  \cdot e_q = p \cdot e_q    =0$. Thus, in
tensor gauge field theory, the VTT amplitude gets non-trivial values in four cases
$M(+2,+1,-2)$,~ $M(-2,+1,+2)$,~ $M(+2,-1,-2)$,  $M(-2,-1,+2)$
and is  explicitly gauge invariant quantity. This completes the analysis of
VTT scattering amplitude in tensor gauge field theory.

\section{\it VTT Amplitude in Superstring  Theory}

In open superstring theory, the {\it linear in momenta} part (\ref{linear}), (\ref{superstringvertex})
of  the full VTT amplitude (\ref{fullcubicvertex}), (\ref{linearcubic})  for
{\it massless vector and  massive tensors} in ten dimensions is given by the
expression\footnote{The details of the calculation are given in section 7 and in Appendix. }
\be\label{superstringamplitude}
M_{string~theory}= \varepsilon_{\alpha\acute{\alpha}}(k_1) e_{\beta}(k_3)
\varepsilon_{\gamma\gamma^{'}}(k_2) ~
F^{\alpha \alpha^{'} \beta \gamma\gamma^{'}}(  k_1,k_3,k_2) ~\delta(k_1 + k_2 +k_3),
\ee
where
\beqa\label{expressionstring}
F^{\alpha \alpha^{'} \beta \gamma\gamma^{'}}  (k_1,k_3,k_2) =
 &+k^{\beta}_{12} ~  (\eta^{\alpha\gamma}\eta^{\alpha^{'}\gamma^{'}} +
\eta^{\alpha\gamma^{'}} \eta^{\alpha^{'}\gamma} ) \nn\\
&+k^{\alpha}_{23}~ (\eta^{\beta\gamma}\eta^{\alpha^{'}\gamma^{'}} +
\eta^{\beta\gamma^{'}} \eta^{\alpha^{'}\gamma} )  \nn\\
&+k^{\alpha^{'}}_{23}~ (\eta^{\beta\gamma}\eta^{\alpha \gamma^{'}} +
\eta^{\beta\gamma^{'}} \eta^{\alpha \gamma} ) \nn\\
&+k^{\gamma}_{31}~ (\eta^{\alpha\beta}\eta^{\alpha^{'} \gamma^{'}} +
\eta^{\alpha^{'}\beta} \eta^{\alpha \gamma^{'}} ) \nn\\
&+k^{\gamma^{'}}_{31}~ (\eta^{\alpha\beta}\eta^{\alpha^{'} \gamma } +
\eta^{\alpha^{'}\beta} \eta^{\alpha \gamma} ) .
\eeqa
Formally comparing these amplitudes in tensor gauge theory (\ref{intymsymmetricvertex})
and in string theory (\ref{expressionstring})  one can see that they
have identical Lorentz structure and are linear in momenta.
But there is a difference in coefficients between these two expressions
in last four terms:
$1/4$ in gauge theory and $1$ in string theory. It seems that this may
contradict to the gauge invariance of both scattering amplitudes.
But one should keep in mind  that in string theory tensor particles are
massive and momentum conservation equation
has no solutions unlike the massless case.
%in this massive case because string amplitude is multiplied by delta function $\delta(k_1 + k_2 +k_3)$ which vanishes.
Nevertheless we shall try to define the amplitude in string theory
by taking a special limit and then to demonstrate that it is also
on mass-shell gauge invariant.

Indeed in string theory tensor states are massive
$
m^{2}_T=  1/\alpha^{'}
$
and the vector boson is massless $m^{2}_{V}=0$, therefore the momentum conservation
$\delta(k_1 + k_2 +k_3)$ {\it has no solutions at all}. The expression for the amplitude
(\ref{superstringamplitude})
has therefore a formal character because
it is multiplied by a delta function which
 vanishes identically. The idea is to  find a reasonable extension of the
string scattering amplitude considering some non-trivial limit that will allow to define it
%a string scattering amplitude
away from the zeros of the
delta function. Let us first consider the wave function of a massive tensor
\beqa\label{polarizationtensors}
&e_{\alpha\alpha^{'}}=
\left(\begin{array}{cccc}
0&0&0&0 \\
0&1&0&0 \\
0&0&-1&0 \\
0&0&0&0 \\
\end{array} \right),
\left(\begin{array}{cccc}
0&0&0&0 \\
0&0&1&0 \\
0&1&0&0 \\
0&0&0&0 \\
\end{array} \right),~~~\epsilon^2 = r^2 + m^2_T  \\
&\left(\begin{array}{cccc}
0&-r/m_T&0&0 \\
-r/m_T&0&0&\epsilon/m_T \\
0&0&0&0 \\
0&\epsilon/m_T&0&0 \\
\end{array} \right),
\left(\begin{array}{cccc}
0&0&-r/m_T&0 \\
0&0&0&0 \\
-r/m_T&0&0&\epsilon/m_T \\
0&0& \epsilon/m_T&0 \\
\end{array} \right),
\left(\begin{array}{cccc}
r^{2}/m_T^2&0&0&- r \epsilon/m^2_T \\
0&-1/2&0&0 \\
0&0&-1/2&0 \\
- r \epsilon/m^2_T&0&0&1 + r^{2}/m^2_T \\
\end{array} \right),\nn
\eeqa
where the first two wave functions correspond to helicities $\pm 2$, the next two
correspond to helicities $\pm 1$ and the last one to $0$.
In the limit $m_T \rightarrow \infty $ and $r \rightarrow 0$
wave functions have a well defined limit and can be used to select
only helicity  $\pm 2$ states as in the tensor gauge field theory. Thus we are
selecting only a subclass of superstring amplitudes with higher
helicities.

Now considering the momentum conservation equation for the choice\footnote{ Without loss
of generality, all momenta and components of the
polarization tensors in the additional six space dimensions
will be taken equal to zero. Note also that discussing only the
properties of massless and massive
modes of strings without loop corrections we could restrict ourselves to lower dimensions.}
$$
k_1  = (m_T, 0, 0, 0,...),~~k_3  =
(\omega, 0, 0, r,...),~~k_2 = (- \sqrt{r^2+m^2_T}, 0, 0, -r,...),
$$
and its complex deformation
$$
k_1  = (m_T, z, iz, 0,...),~~k_3  =
(\omega, -z, -iz, r,...),~~k_2 = (- \sqrt{r^2+m^2_T}, 0, 0, -r,...)
$$
one can see that the equation fulfills  if $2 m_T r =0 $.
Therefore, if we take the limit  $m_T \rightarrow \infty$,
$r \rightarrow 0$, so that $ m_T r \rightarrow 0 $,~
then the momentum conservation indeed
can be fulfilled. Thus it seems possible to define this amplitude in superstring model.
Physically, this limit corresponds to the interaction
between infinitely heavy tensor bosons  $(m_T \rightarrow \infty)$ and massless
vector bosons which are in the deep infrared region of the spectrum $(r  \rightarrow 0 )$.
It is like an exchange interaction between very heavy ions at rest and photons of tiny energy.

Thus in {\it superstring  theory} the amplitude (\ref{superstringamplitude}),
(\ref{expressionstring}) can be written in the form
\beqa\label{ymsymmetricvertex2}
&M_{string~theory}=\varepsilon_{\alpha\acute{\alpha}}(k) e_{\beta}(p)  \varepsilon_{\gamma\gamma^{'}}(q)
F^{\alpha\acute{\alpha}\beta\gamma\acute{\gamma}}(k,p,q) = \nn\\  &   ~
=4 (k  \cdot e_p)  ~~(\varepsilon_k  \cdot \varepsilon_q)
 - 8 (p  \cdot \varepsilon_k \cdot \varepsilon_q \cdot e_p)
 +   8 (e_p  \cdot \varepsilon_k \cdot \varepsilon_q \cdot p).
\eeqa
The momenta we shall take as above:
\be\label{deformation2}
k  = (m_T, z, iz, 0,...),~~p  =
(\omega, -z, -iz, r,...),~~q = (-\sqrt{r^2 + m^2_T}, 0, 0, - r,...),
\ee
$(\omega^2 = r^2)$ and the polarization tensors can be chosen as follows
$$
e_{k}={1\over \sqrt{2}} ({2z \over m_T}, 1, -i, -{2z \over m_T} ,...),~~
e_p={1\over \sqrt{2}}(-{z \over \omega}, 1, -i, {z \over r} ,...),~~
e_q={1\over \sqrt{2}}(0, 1,  i ,0,... )
$$
where $\varepsilon_{\alpha\acute{\alpha}}(k)= e_{\alpha}(k) e_{\acute{\alpha}}(k)$,~
$\varepsilon_{\gamma\acute{\gamma}}(q)= e_{\gamma}(q) e_{\acute{\gamma}}(q)$.
We can calculate now the amplitude  (\ref{ymsymmetricvertex2}) using the
relations $e_k \cdot e_p=0,~e_k \cdot e_q =  1,~ e_p \cdot e_q= 1$,~
$k \cdot e_p=  z(2+m_T/r)/\sqrt{2}$, ~ $p \cdot e_k = - z(2+ 4 r/m_T)/\sqrt{2}$.
Thus we shall get nontrivial analytical continuation of the VTT amplitude in string theory
\beqa
&M(+2,+1,-2)_{string~theory}=4 (k  \cdot e_p)  ~~(\varepsilon_k  \cdot \varepsilon_q)
 - 8 (p  \cdot \varepsilon_k \cdot \varepsilon_q \cdot e_p)+\nn\\
& +   8 (e_p  \cdot \varepsilon_k \cdot \varepsilon_q \cdot p)=
 12\sqrt{2} z
 + 2\sqrt{2} z({m_T\over r} + 8{ r \over m_T }),
\eeqa
which has a part which is identical with the massless tensor gauge theory $12\sqrt{2} z$
in (\ref{ymantisymmetricvertexspinorial1}) and an additional part which depends on the
mass of the tensor particle $m_T$. One should take now the limit
 $m_T \rightarrow \infty$,
$r \rightarrow 0$,~ $ m_T r \rightarrow 0 $,~ keeping $z {m_T\over r}=Z$ fixed, so that
$M(+2,+1,-2)_{string}=2\sqrt{2} Z$.

\section{\it Mass-shell gauge invariance in superstring  theory}

This expression is also {\it on mass-shell gauge invariant}
\beqa\label{variationb}
&e_{\beta}(p) \rightarrow e_{\beta}(p) +  \xi ~ p_{\beta}\nn\\
&\varepsilon_{\alpha \alpha^{'} }(k) \rightarrow \varepsilon_{\alpha  \alpha^{'} }(k)
+ k_{\alpha} \xi_{\alpha^{'}} + k_{\alpha^{'}} \xi_{\alpha},~~~~~~k^2=-m^2_T,~~~~k \cdot \xi =0
\eeqa
where $\xi$ and $\xi_{\alpha}$ are gauge parameters.
The gauge variation  of (\ref{ymsymmetricvertex2}) under (\ref{variationb}) $\delta e_p \sim p$ is
\be
\delta M_{string~theory} = 4 (k \cdot p) ~ (\varepsilon_k \cdot \varepsilon_q)
- 8 (p  \cdot \varepsilon_k \cdot \varepsilon_q \cdot p)
 +  8  (p  \cdot \varepsilon_k \cdot \varepsilon_q \cdot p) =  4 r m_T \rightarrow 0
\ee
and under (\ref{variationb}) $\delta \varepsilon_k \sim k \otimes \xi + \xi \otimes k$ is
\beqa
&\delta M_{string~theory} = 8 (k \cdot e_p) ~  (k \cdot \varepsilon_q \cdot \xi)-\\
&- 8(p  \cdot  k) ~ (\xi \cdot \varepsilon_q \cdot e_p)
- 8(p  \cdot \xi) ~ (k  \cdot \varepsilon_q \cdot e_p)
 +   8 (e_p  \cdot k) ~(\xi \cdot \varepsilon_q \cdot p)
 +   8 (e_p  \cdot  \xi) ~(k \cdot \varepsilon_q \cdot p)=\nn\\
& = -8 r m_T~ (\xi \cdot \varepsilon_q \cdot e_p) \rightarrow 0, \nn
\eeqa
because $p  \cdot  k = r m_T \rightarrow 0,~~  k  \cdot e_q =0,~ p \cdot e_q    =0$.
Thus, this expression is also gauge invariant on mass-shell for gauge variations
$\xi_\alpha$ in any direction of the ten dimensional space-time.

\section{\it Effective Action in Terms of Tensor Gauge Fields}

In the {\it full superstring   amplitude} (\ref{fullcubicvertex}), (\ref{linearcubic})  together with
the linear part (\ref{linear}) we have
also  an additional term which is cubic $k^3$ in momenta
\beqa\label{highderivative}
   \alpha^{'}    \{~
&-&\eta^{\alpha^{'} \gamma^{'}} k^{\alpha}_3 k^{\gamma}_1 k^{\beta}_2
+\eta^{\beta \gamma^{'}} k^{\alpha}_3 k^{\alpha^{'}}_2 k^{\gamma}_1
- \eta^{\beta\alpha^{'}} k^{\alpha }_2   k^{\gamma}_3   k^{\gamma^{'}}_1~ +\nn \\
&+&(k_2\cdot k_3)~ k^{\alpha^{'}}_2(\eta^{\alpha \gamma^{'}}\eta^{\beta\gamma}
- \eta^{\alpha \gamma}\eta^{\beta \gamma^{'}}  ) +
 (k_1\cdot k_3) k^{\gamma^{'}}_3  (\eta^{\alpha \beta}\eta^{\alpha^{'}\gamma}
- \eta^{\alpha \gamma}\eta^{\alpha^{'}\beta }  )+\nn\\
&+&(k_1 \cdot k_2)[~
(\eta^{\beta\gamma} \eta^{\alpha^{'}\gamma^{'} }
                             - \eta^{\alpha^{'} \gamma} \eta^{\beta\gamma^{'}})k^{\alpha }_{3}
 + (\eta^{\alpha \gamma} \eta^{\beta\gamma^{'} }
                             - \eta^{\gamma\beta} \eta^{\alpha \gamma^{'}})k^{\alpha^{'}}_{3}
\nn\\
&&~~~~~~~~~~~~~~~~~ +(\eta^{\alpha\gamma^{'} }\eta^{\beta \alpha^{'}}
                             - \eta^{\alpha \beta} \eta^{\alpha^{'}\gamma^{'}})k^{\gamma }_{3}
+ (\eta^{\alpha \beta} \eta^{\alpha^{'} \gamma }
                             - \eta^{\alpha \gamma}\eta^{\alpha^{'}\beta} )k^{\gamma^{'}}_{3}
\nn\\
&&~~~~~~~~~~~~~~~~~~+(\eta^{\alpha \gamma^{'}} \eta^{\alpha^{'} \gamma}
                             - \eta^{\alpha \gamma} \eta^{\alpha^{'}\gamma^{'} })
                             k^{\beta}_{1}~]~\}.
\eeqa
In particular it contains  scalar products $ k_1\cdot k_2 $,~ $k_1 \cdot k_3$ and
$k_2 \cdot k_3$.
In the three particle scattering amplitude, which we are considering here, they can take
only  fixed values $
k_1 \cdot k_3 = k_2 \cdot k_3 = 0,~~~2 \alpha^{'} k_1 \cdot k_2 =
- \alpha^{'} (k^2_1+k^2_2) = 2,
$
therefore they do not appear in the final expression (\ref{linearcubic}).
Nevertheless let us keep them all, as they are, in order to examine if
they can be reproduced by an effective Lagrangian which
is constructed using generalized field strength tensors (\ref{tensors}).

Naturally we should try to associate these cubic terms with a gauge invariant effective
Lagrangian which has higher derivatives. Indeed, there are two independent gauge
invariant forms which can be constructed in tensor gauge field theory using the field
strengths (\ref{tensors}) and are cubic in derivatives
\be\label{effectiveaction1}
\CL_{\partial}=\alpha^{'} [~Tr(G_{\mu\nu,\lambda} G_{\nu \rho } G_{\rho \mu,\lambda})
+{1 \over 2} Tr(G_{\mu\nu} G_{\nu \rho,\lambda\lambda } G_{\rho\mu})~]
\ee
and
\beqa\label{effectiveaction2}
 \CL^{'}_{\partial}=\alpha^{'} [~&- Tr(G_{\mu\nu,\lambda} G_{\nu \rho } G_{\rho \mu,\lambda})
+ Tr(G_{\mu\lambda,\lambda} G_{\mu\nu } G_{\nu \rho,\rho})
+Tr(G_{\mu\nu,\lambda} G_{\mu \rho } G_{\rho \lambda,\nu})~+\nn\\
&+Tr(G_{\mu\nu} G_{\mu \rho,\nu}  G_{\rho \lambda,\lambda})+
Tr(G_{\mu\nu} G_{\mu \rho,\lambda}  G_{\rho \lambda,\nu})+
Tr(G_{\mu\nu} G_{\mu \rho,\lambda}  G_{\rho\nu, \lambda})+\nn\\
&+Tr(G_{\mu\lambda,\lambda} G_{\mu \nu,\rho}  G_{\nu\rho })+
Tr(G_{\mu\nu,\lambda} G_{\mu \rho,\lambda}  G_{\rho\nu })+
Tr(G_{\mu\nu,\lambda} G_{\mu \rho,\nu}  G_{\rho \lambda})+\nn\\
&+2Tr(G_{\mu\nu } G_{\nu\lambda, \rho \mu} G_{\rho\lambda})-
Tr(G_{\mu\nu} G_{\nu\rho, \lambda\lambda} G_{\rho \mu})~]
\eeqa
It is interesting
that  reproducing the higher derivative part (\ref{highderivative}) of the VTT vertex,
there are no ``traces" of any higher derivative string (gravity) vertex VVT
between two vectors and a tensor (two photons and a graviton).
What is also striking is that one reproduces all terms with scalar products of momenta
$ k_1\cdot k_2 $,~ $k_1 \cdot k_3$ and $k_2 \cdot k_3$ in (\ref{highderivative}).
In our on-mass-shell  scattering amplitude they have fixed values and did not ``show up", but they will
certainly contribute to other more complicated amplitudes. Therefore it is
important that they are present in the effective Lagrangian.

In the next section we shall present the actual calculation of the
superstring scattering amplitude (\ref{superstringamplitude}), (\ref{fullcubicvertex}).
As our calculation shows, the $(\alpha^{'})^4 k^5$ terms are absent in superstring theory.

\section{{\it Open Type I Superstring Tree-Level Amplitudes}}

To set up notation let us begin with the simplest example of the tree-level
scattering amplitude for three on-shell massless vector bosons.
The vertex operator has the following form\footnote{In eq.~(\ref{vectorsvertex}) and below, the superscript -1 stands for the $(-1)$-ghost picture.}
\cite{Green:1987sp,Polchinski:1998rq}:
\beqa\label{vectorsvertex}
&\CV^0=e_{\alpha}(k) (\dot{X}^{\alpha}
-2i \alpha^{'} k\cdot \psi \psi^{\alpha} ) e^{ik  X}(y) \nn\\
&\CV^{-1}=e_{\alpha}(k)  e^{-\phi} \psi^{\alpha}   e^{ik  X}(y)
\eeqa
and we shall represent the disk as the upper half-plane so that the boundary
coordinate $y$ is real $y \in [-\infty,+\infty]$. The tree amplitude can take the form
\beqa
&\CV^{\mu_1\mu_2\mu_3}_{a_{1} a_{2} a_{3} } (k_1 ,k_2 ,k_3)
= F^{\mu_1\mu_2\mu_3}(k_1 ,k_2 ,k_3)
tr(\lambda^{a_{1}} \lambda^{a_{2}}\lambda^{a_{3}}) +
 F^{\mu_2\mu_1\mu_3}(k_2 ,k_1 ,k_3)
tr(\lambda^{a_{2}} \lambda^{a_{1}} \lambda^{a_{3}})\nn\, ,
\eeqa
where $\lambda^{a}$ are isotopic matrices and the matrix element $F$ is given below
\beqa
& F^{\mu_1\mu_2\mu_3}(k_1 ,k_2 ,k_3) = <c \CV^{-1}(y_1) ~c \CV^{-1}(y_2)~ c \CV^{0}(y_3) =\nn\\
&=< ce^{-\phi}\psi^{\mu_1}e^{ik_1 X}(y_1)~ce^{-\phi}\psi^{\mu_2}e^{ik_2 X}(y_2) ~
c (\dot{X}^{\mu_3} -2i \alpha^{'} k_3\cdot\psi \psi^{\mu_3} ) e^{ik_3 X}(y_3) >  \nn\\
& =~ y_{12}~ y_{23}~ y_{13}~y^{-1}_{12}  \nn\\
&\{ F^{\mu_3}_{y_3}~\eta^{\mu_1\mu_2}~ y^{-1}_{12}
+ 2i \alpha^{'} ~k^{\mu_1}_3 y^{-1}_{13}~\eta^{\mu_2\mu_3} y^{-1}_{23}
- 2i \alpha^{'}~ k^{\mu_2}_3 y^{-1}_{23}~\eta^{\mu_1\mu_3} y^{-1}_{13}\}. \nn
\eeqa
Here $y_{ij}=y_i -y_j$,~ $ y_3 < y_2 < y_1  $ and we have to sum over two orderings of the vertex operators on the disk.
The vector function $F^{\mu_3}_{y_3}$ is given by the expression
\beqa\label{f3}
&F^{\mu_3}_{y_3}= -2 i \alpha^{'}~({k^{\mu_3}_1 \over y_3-y_1}
+{k^{\mu_3}_2 \over y_3-y_2})=-2 i \alpha^{'}~{k^{\mu_3}_1 y_{12}\over y_{13}y_{23}}.\nn
\eeqa
The relevant correlation functions are
\beqa
<c(y_1)c(y_2)c(y_3)> = y_{12}~ y_{23}~ y_{13},~<e^{-\phi}(y_1)e^{-\phi}(y_2)>
 = y^{-1}_{12}\,, \nn
\eeqa
while the contraction of  world-sheet fermions is
\beqa
 <\psi^{\mu}(y_1)\psi^{\nu}(y_2)> =\eta^{\mu \nu }~ y^{-1}_{12}.\nn
\eeqa
All bosons are on mass-shell
$
\alpha^{'} k^{2}_{1}=\alpha^{'} k^{2}_{2}=\alpha^{'} k^{2}_{3}=0
$
and $k_1 +k_2 +k_3 =0$. Their wave functions are
$
e_{\mu_1}(k_1),~~~e_{\mu_2}(k_2),~~~e_{\mu_3}(k_3)
$
and are transversal to the corresponding momenta $k_{i} \cdot e(k_i)=0,~ i=1,2,3$.
One sees that the matrix element $F^{\mu_1\mu_2\mu_3}$ is linear in momentum
\be
[F^{\mu_1\mu_2\mu_3}] \sim
  \alpha^{'}   k\, .
\ee
Unlike the bosonic
open string amplitude, there is no $k^3$ term and so no $G^3$
term in the low energy effective action. Thus for the $F^{\mu_1\mu_2\mu_3}(k_1 ,k_2 ,k_3)
tr(\lambda^{a_{1}} \lambda^{a_{2}}\lambda^{a_{3}})$ we have
\beqa
& 2i\alpha^{'}[k^{\mu_1}_{3}\eta^{\mu_2\mu_3}- k^{\mu_2}_{3}\eta^{\mu_1\mu_3}
-k^{\mu_3}_{1} \eta^{\mu_1\mu_2} ]~~
tr(\lambda^{a_{1}} \lambda^{a_{2}}\lambda^{a_{3}}).
\eeqa
Adding the equal term\footnote{One should use momentum conservation and
transversality of the wave functions.}
$
2 i \alpha^{'}
[-k^{\mu_1}_{2}\eta^{\mu_2\mu_3} + k^{\mu_2}_{1}\eta^{\mu_3\mu_1}+
k^{\mu_3}_{2}\eta^{\mu_1\mu_2} ]~~
tr(\lambda^{a_{1}} \lambda^{a_{2}}\lambda^{a_{3}})
$
and  the
reversed cyclic orientation amplitude
$a_1,\mu_1,k_1 \leftrightarrow a_2,\mu_2, k_2$, we can get the total matrix element:
\beqa
&i \alpha^{'}
[~(k_3-k_{2})^{\mu_1}\eta^{\mu_2\mu_3} +  (k_1-k_3)^{\mu_2}\eta^{\mu_3\mu_1} +(k_2-
k_1)^{\mu_3}\eta^{\mu_1\mu_2} ]~~
 tr([\lambda^{a_{1}} , \lambda^{a_{2}}]\lambda^{a_{3}}).
\eeqa
This expression coincides with the Yang-Mills vertex  projected to the mass-shell.

\subsection{\it Tree-Level Amplitude for Two Symmetric Tensors and a Vector}

The vertex operator for the symmetric rank-2 tensor boson $T_S$ on the second level is
\be\label{symmetricvertex}
\CV^{-1} =\varepsilon_{\alpha\alpha^{'}}(k)  e^{-\phi} \psi^{\alpha}(\dot{X}^{\alpha^{'}}
-2i \alpha^{'} k\cdot \psi \psi^{\alpha^{'}} ) e^{ik  X}(y)
\ee
and together with the vertex (\ref{vectorsvertex}) can be used to calculate now the scattering
amplitude between a vector and two tensor bosons:
\beqa\label{vecror2tensorvertex}
&\CV^{\alpha \alpha^{'} \beta \gamma\gamma^{'}}_{a  b c } (k  ,p ,q)
= F^{\alpha \alpha^{'} \beta \gamma\gamma^{'}}  (k  ,p ,q)~
tr(\lambda^{a } \lambda^{b}\lambda^{c}) +
 F^{\gamma\gamma^{'} \beta \alpha \alpha^{'}}  (q  ,p ,k)~
tr(\lambda^{c} \lambda^{b} \lambda^{a})
\eeqa
where their wave functions are:
\be
\varepsilon_{\alpha \alpha^{'}}(k),~~~e_{ \beta}(p),~~~\varepsilon_{\gamma\gamma^{'}}(q).
\ee
We shall define for convenience
$k_1 \equiv k, k_2 \equiv q, k_3 \equiv p,$ and  $k_1 +k_2 +k_3 =0$.
The mass-shell conditions are
\be
\alpha^{'} k^{2}_{1}=\alpha^{'} k^{2}_{2}= -1,~   \alpha^{'} k^{2}_{3}=0
\ee
and therefore it follows that
\be\label{massshell}
k_1 \cdot k_3 = k_2 \cdot k_3 = 0,~~~2 \alpha^{'} k_1 \cdot k_2 =
- \alpha^{'} (k^2_1+k^2_2) = 2.
\ee
We have to calculate the correlation function:
\beqa
&<:c e^{-\phi} \psi^{\alpha}(\dot{X}^{\alpha^{'}}
-2i \alpha^{'} k\cdot \psi \psi^{\alpha^{'}} ) e^{ik  X}(y_1):~
:c e^{-\phi} \psi^{\gamma}(\dot{X}^{\gamma^{'}}
-2i \alpha^{'} q\cdot \psi \psi^{\gamma^{'}} ) e^{iq X}(y_2): ~\nn\\
&:c  (\dot{X}^{\beta}
-2i \alpha^{'} p\cdot \psi \psi^{\beta} )e^{ip X}(y_3): >
\eeqa
We shall split it into four terms. The first one gives (the details
of the calculation are given in the
Appendix)
\beqa
& <c e^{-\phi} \psi^{\alpha} \dot{X}^{\alpha^{'}}  e^{ik_1  X}(y_1)~
c e^{-\phi} \psi^{\gamma} \dot{X}^{\gamma^{'}}  e^{ik_2 X}(y_2)~
c (\dot{X}^{\beta}
-2i \alpha^{'} k_3\cdot \psi \psi^{\beta} )e^{ik_3 X}(y_3) >= \nn\\
&=i(2 \alpha^{'})^2 [\eta^{\alpha \gamma } (\eta^{\beta\gamma^{'} } k^{\alpha^{'}}_{2}
+ \eta^{\alpha^{'}\beta } k^{\gamma^{'}}_{3} +\eta^{\alpha^{'}\gamma^{'} } k^{\beta}_{1})
+ \eta^{\alpha^{'}\gamma^{'} } (\eta^{\alpha \beta } k^{\gamma }_{3}
+ \eta^{\beta \gamma} k^{\alpha}_{2} )]\nn\\
&(-2 i\alpha^{'})^3 [\eta^{\alpha \gamma } k^{\beta}_{1}
+\eta^{\alpha \beta } k^{\gamma}_{3} +\eta^{ \gamma\beta } k^{\alpha}_{2}]
k^{\alpha^{'}}_{2} k^{\gamma^{'}}_{3}  \nn
\eeqa
and contains {\it  linear as well as cubic in momentum expressions}. The other three remaining terms
have only cubic in momentum expressions. Indeed the second one gives
\beqa
& <c e^{-\phi} \psi^{\alpha} \dot{X}^{\alpha^{'}}  e^{ik_1  X}(y_1)~
c e^{-\phi} \psi^{\gamma} (-2i \alpha^{'}) k_2 \cdot \psi \psi^{\gamma^{'}}
e^{ik_2 X}(y_2)~ c (\dot{X}^{\beta}
-2i \alpha^{'} k_3\cdot \psi \psi^{\beta} )e^{ik_3 X}(y_3) >= \nn\\
&=(-2 i \alpha^{'})^3 k^{\alpha^{'}}_2 [\eta^{\alpha \gamma} k^{\beta}_2 k^{\gamma^{'}}_3
+\eta^{\beta \gamma^{'}} k^{\alpha}_2 k^{\gamma}_3
- \eta^{\beta\gamma} k^{\alpha }_2 k^{\gamma^{'}}_3
-\eta^{\alpha \gamma^{'}} k^{\beta}_2 k^{\gamma}_3
+ k_2\cdot k_3(\eta^{\alpha \gamma^{'}}\eta^{\beta\gamma}
- \eta^{\alpha \gamma}\eta^{\beta \gamma^{'}}  )]\nn
\eeqa
and contains {\it only cubic in momentum terms}.  A new feature of this expression
is that it contains a scalar product $(k_2\cdot k_3) $.
The third  one gives
\beqa
& <c e^{-\phi} \psi^{\alpha} (-2i \alpha^{'}) k_1 \cdot \psi \psi^{\alpha^{'}} e^{ik_1  X}(y_1)~
c e^{-\phi} \psi^{\gamma} \dot{X}^{\gamma^{'}}  e^{ik_2 X}(y_2)~
c (\dot{X}^{\beta}-2i \alpha^{'} k_3\cdot \psi \psi^{\beta} )e^{ik_3 X}(y_3) >= \nn\\
&=(-2 i \alpha^{'})^3 k^{\gamma^{'}}_3 [\eta^{\alpha \gamma} k^{\beta}_1 k^{\alpha^{'}}_3
-\eta^{\alpha\beta } k^{\alpha^{'}}_3 k^{\gamma}_1
+ \eta^{\beta\alpha^{'}} k^{\alpha }_3 k^{\gamma}_1
-\eta^{\alpha^{'} \gamma} k^{\beta}_1 k^{\alpha}_3
+ k_1\cdot k_3(\eta^{\alpha \beta}\eta^{\alpha^{'}\gamma}
- \eta^{\alpha \gamma}\eta^{\alpha^{'}\beta }  )]\nn
\eeqa
and also contains {\it only  cubic in momentum expressions } as well as
a scalar product$(k_1\cdot k_3)$. Finally, the last term gives
\beqa
 <&c& e^{-\phi} \psi^{\alpha} (-2i \alpha^{'}) k_1 \cdot \psi \psi^{\alpha^{'}} e^{ik_1  X}(y_1)~
c e^{-\phi} \psi^{\gamma} (-2i \alpha^{'}) k_2 \cdot \psi \psi^{\gamma^{'}}~
c (\dot{X}^{\beta}-2i \alpha^{'} k_3\cdot \psi \psi^{\beta} )e^{ik_3 X}(y_3) > \nn\\
&&~~~~~~~~~~~~~~~~~~~~~~~~~~~~~~ =(-2 i \alpha^{'})^3
\nn \\
\{&+& k^{\beta }_{1}
 [  \eta^{\alpha \gamma}   (- k_1\cdot k_2 \eta^{\alpha^{'}\gamma^{'} }
  + k^{\gamma^{'}}_1 k^{\alpha^{'}}_2  )
- k^{\alpha}_2 (- k^{\gamma}_1  \eta^{\alpha^{'} \gamma^{'} }
 +  k^{\gamma^{'}}_1 \eta^{\alpha^{'} \gamma  }  )
 + \eta^{\alpha \gamma^{'}}  (-   k^{\gamma}_1 k^{\alpha^{'}}_2
 + k_1\cdot k_2 \eta^{\alpha^{'}\gamma }  )]
 \nn\\
&-& k^{\alpha}_3 [  \eta^{\beta \gamma}
(- k_1\cdot k_2 \eta^{\alpha^{'}\gamma^{'} }
 + k^{\gamma^{'}}_1 k^{\alpha^{'}}_2  )
-   k^{\beta}_2 (- k^{\gamma}_1  \eta^{\alpha^{'} \gamma^{'} }
 +  k^{\gamma^{'}}_1 \eta^{\alpha^{'} \gamma  } )+
   \eta^{\beta \gamma^{'}}   (-  k^{\gamma}_1 k^{\alpha^{'}}_2
 + k_1\cdot k_2 \eta^{\alpha^{'}\gamma }  )]
\nn\\
&-& k^{\alpha^{'}}_3   [ \eta^{\beta \gamma}
( k_1\cdot k_2 \eta^{\alpha\gamma^{'} }
- k^{\gamma^{'}}_1 k^{\alpha}_2  )
-   k^{\beta}_2  (- k^{\gamma^{'}}_1  \eta^{\alpha \gamma }
 +  k^{\gamma}_1 \eta^{\alpha \gamma^{'}  }   )+
   \eta^{\beta \gamma^{'}}   (   k^{\gamma}_1 k^{\alpha}_2
 - k_1\cdot k_2 \eta^{\alpha \gamma }   )]
\nn\\
&+& k^{\gamma}_3   [  \eta^{\alpha\beta }
(- k_1\cdot k_2 \eta^{\alpha^{'}\gamma^{'} }
 + k^{\gamma^{'}}_1 k^{\alpha^{'}}_2  )
-   k^{\beta}_1   (- k^{\alpha}_2  \eta^{\alpha^{'} \gamma^{'} }
 +  k^{\alpha^{'}}_2 \eta^{\alpha \gamma^{'}  }  )+
   \eta^{\beta \alpha^{'}}  (-   k^{\gamma^{'}}_1 k^{\alpha}_2
  + k_1\cdot k_2 \eta^{\alpha\gamma^{'} } )]
\nn\\
&+& k^{\gamma^{'}}_3  [  \eta^{\alpha\beta }
( k_1\cdot k_2 \eta^{\alpha^{'}\gamma }
  - k^{\gamma}_1 k^{\alpha^{'}}_2  )
-   k^{\beta}_1   ( k^{\alpha}_2  \eta^{\alpha^{'} \gamma}
  -  k^{\alpha^{'}}_2 \eta^{\alpha \gamma  }   )+
   \eta^{\beta \alpha^{'}}   (  k^{\gamma}_1 k^{\alpha}_2
  - k_1\cdot k_2 \eta^{\alpha\gamma } )]
\nn\\
&+& k_1 \cdot k_3  [  \eta^{\beta \gamma}
(  k^{\alpha^{'}}_2  \eta^{\alpha\gamma^{'}}_1
-  k^{\alpha}_2  \eta^{\alpha^{'}\gamma^{'}}  )
-   k^{\beta}_2  (-  \eta^{\alpha \gamma } \eta^{\alpha^{'}\gamma^{'}}
  +  \eta^{\alpha^{'}\gamma} \eta^{\alpha \gamma^{'}  }  )+
   \eta^{\beta \gamma^{'}}   (  k^{\alpha}_2 \eta^{\alpha^{'} \gamma }
 - k^{\alpha^{'}}_2 \eta^{\alpha \gamma }   )]
\nn\\
&-& k_2 \cdot k_3   [  \eta^{\alpha\beta }
(  k^{\gamma^{'}}_1  \eta^{\alpha^{'}\gamma}_1
-  k^{\gamma}_1  \eta^{\alpha^{'}\gamma^{'}}  )
-   k^{\beta}_1   (- \eta^{\alpha \gamma } \eta^{\alpha^{'}\gamma^{'}}
 + \eta^{\alpha^{'}\gamma} \eta^{\alpha \gamma^{'}  }  ) +
   \eta^{\beta \alpha^{'}}   (   k^{\gamma}_1 \eta^{\alpha \gamma^{'} }
 - k^{\gamma^{'}}_1 \eta^{\alpha \gamma }  )]
\nn
\}
\eeqa
and again contains {\it  only cubic in momentum expressions} as well as a scalar product
$(k_1\cdot k_2)$. The scalar products $k_1 \cdot k_3$ and $k_2 \cdot k_3$ can be dropped because
of (\ref{massshell}).
Summing all remaining terms together one can see that many terms which
are cubic in momentum cancel each
other so that we are left with the expression
\beqa\label{fullcubicvertex}
& i(2 \alpha^{'})^2 [~\eta^{\alpha \gamma } (\eta^{\beta\gamma^{'} } k^{\alpha^{'}}_{2}
+ \eta^{\alpha^{'}\beta } k^{\gamma^{'}}_{3} +\eta^{\alpha^{'}\gamma^{'} } k^{\beta}_{1})
+ \eta^{\alpha^{'}\gamma^{'} } (\eta^{\alpha \beta } k^{\gamma }_{3}
+ \eta^{\beta \gamma} k^{\alpha}_{2} )~]+
\nn\\
&(-2 i \alpha^{'})^3  \{~
-\eta^{\alpha^{'} \gamma^{'}} k^{\alpha}_3 k^{\gamma}_1 k^{\beta}_2
+\eta^{\beta \gamma^{'}} k^{\alpha}_3 k^{\alpha^{'}}_2 k^{\gamma}_1
- \eta^{\beta\alpha^{'}} k^{\alpha }_2   k^{\gamma}_3   k^{\gamma^{'}}_1~]+~~~~~~~~~~~~~~~~~~~
\nn\\
& (k_1 \cdot k_2)[~
+ (\eta^{\beta\gamma} \eta^{\alpha^{'}\gamma^{'} }
                             - \eta^{\alpha^{'} \gamma} \eta^{\beta\gamma^{'}})k^{\alpha }_{3}
 + (\eta^{\alpha \gamma} \eta^{\beta\gamma^{'} }
                             - \eta^{\gamma\beta} \eta^{\alpha \gamma^{'}})k^{\alpha^{'}}_{3}
\nn\\
&~~~~~~~~~~~~~+ (\eta^{\alpha\gamma^{'} }\eta^{\beta \alpha^{'}}
                             - \eta^{\alpha \beta} \eta^{\alpha^{'}\gamma^{'}})k^{\gamma }_{3}
+ (\eta^{\alpha \beta} \eta^{\alpha^{'} \gamma }
                             - \eta^{\alpha \gamma}\eta^{\alpha^{'}\beta} )k^{\gamma^{'}}_{3}
\nn\\
&~~~~~~~+(\eta^{\alpha \gamma^{'}} \eta^{\alpha^{'} \gamma}
                             - \eta^{\alpha \gamma}
                             \eta^{\alpha^{'}\gamma^{'} })k^{\beta}_{1}~]~\}.
\eeqa
Note that the terms proportional to the nonzero product
$2 \alpha^{'}k_1 \cdot k_2 =2$  (\ref{massshell}) can also be dropped because
they are antisymmetric with respect to the indices $\alpha \alpha^{'}$ and
$\gamma \gamma^{'}$ while the wave functions  $\varepsilon_{\alpha\alpha^{'}}$
and $\varepsilon_{\gamma\gamma^{'}}$ are symmetric.

Thus, we arrive to the
following expression
\beqa\label{linearcubic}
& i(2 \alpha^{'})^2 [~\eta^{\alpha \gamma } (\eta^{\beta\gamma^{'} } k^{\alpha^{'}}_{2}
+ \eta^{\alpha^{'}\beta } k^{\gamma^{'}}_{3} +\eta^{\alpha^{'}\gamma^{'} } k^{\beta}_{1})
+ \eta^{\alpha^{'}\gamma^{'} } (\eta^{\alpha \beta } k^{\gamma }_{3}
+ \eta^{\beta \gamma} k^{\alpha}_{2} )~]+
\nn\\
&(-2 i \alpha^{'})^3  [~
-\eta^{\alpha^{'} \gamma^{'}} k^{\alpha}_3 k^{\gamma}_1 k^{\beta}_2
+\eta^{\beta \gamma^{'}} k^{\alpha}_3 k^{\alpha^{'}}_2 k^{\gamma}_1
- \eta^{\beta\alpha^{'}} k^{\alpha }_2   k^{\gamma}_3   k^{\gamma^{'}}_1~],
\eeqa
which is {\it linear and cubic in momenta}. Its linear in momenta part is
\beqa\label{linear}
& 4[\eta^{\alpha \gamma } (\eta^{\beta\gamma^{'} } q^{\alpha^{'}}
+ \eta^{\alpha^{'}\beta } p^{\gamma^{'}}  +\eta^{\alpha^{'}\gamma^{'} } k^{\beta} )
+ \eta^{\alpha^{'}\gamma^{'} } (\eta^{\alpha \beta } p^{\gamma }
+ \eta^{\beta \gamma} q^{\alpha}  )].
\eeqa
It should be symmetrized over $\alpha \alpha^{'}$ and $\gamma\gamma^{'}$
\beqa
& +(\eta^{\alpha \gamma }  \eta^{\alpha^{'}\gamma^{'} }
+  \eta^{\alpha\gamma^{'}  } \eta^{\alpha^{'}\gamma }) k^{\beta}\nn\\
&+(\eta^{\beta \gamma }  \eta^{\alpha^{'}\gamma^{'} }
+  \eta^{\beta\gamma^{'}  } \eta^{\alpha^{'}\gamma })q^{\alpha} \nn\\
&+(\eta^{\beta \gamma }  \eta^{\alpha\gamma^{'} }
+  \eta^{\beta\gamma^{'}  } \eta^{\alpha \gamma })q^{\alpha^{'}} \nn\\
&+(\eta^{\alpha \beta }  \eta^{\alpha^{'}\gamma^{'} }
+  \eta^{\alpha^{'}\beta  } \eta^{\alpha\gamma^{'} }) p^{\gamma}\nn\\
&+(\eta^{\alpha \beta }  \eta^{\alpha^{'}\gamma }
+  \eta^{\alpha^{'}\beta  } \eta^{\alpha\gamma }) p^{\gamma^{'}}
\eeqa
and we can add to it an equal term
\beqa
& -(\eta^{\alpha \gamma }  \eta^{\alpha^{'}\gamma^{'} }
+  \eta^{\alpha\gamma^{'}  } \eta^{\alpha^{'}\gamma }) q^{\beta}\nn\\
&-(\eta^{\beta \gamma }  \eta^{\alpha^{'}\gamma^{'} }
+  \eta^{\beta\gamma^{'}  } \eta^{\alpha^{'}\gamma })p^{\alpha} \nn\\
&-(\eta^{\beta \gamma }  \eta^{\alpha\gamma^{'} }
+  \eta^{\beta\gamma^{'}  } \eta^{\alpha \gamma })p^{\alpha^{'}} \nn\\
&-(\eta^{\alpha \beta }  \eta^{\alpha^{'}\gamma^{'} }
+  \eta^{\alpha^{'}\beta  } \eta^{\alpha\gamma^{'} }) k^{\gamma}\nn\\
&-(\eta^{\alpha \beta }  \eta^{\alpha^{'}\gamma }
+  \eta^{\alpha^{'}\beta  } \eta^{\alpha\gamma }) k^{\gamma^{'}}
\eeqa
in order to get a symmetric expression:
\beqa\label{superstringvertex}
^{(1)}F^{\alpha \alpha^{'} \beta \gamma\gamma^{'}} (k  ,p ,q) =
&+(k-q)^{\beta}  (\eta^{\alpha\gamma}\eta^{\alpha^{'}\gamma^{'}} +
\eta^{\alpha\gamma^{'}} \eta^{\alpha^{'}\gamma} ) \nn\\
&+(q-p)^{\alpha} (\eta^{\beta\gamma}\eta^{\alpha^{'}\gamma^{'}} +
\eta^{\beta\gamma^{'}} \eta^{\alpha^{'}\gamma} )  \nn\\
&+(q-p)^{\alpha^{'}} (\eta^{\beta\gamma}\eta^{\alpha \gamma^{'}} +
\eta^{\beta\gamma^{'}} \eta^{\alpha \gamma} ) \nn\\
&+(p-k)^{\gamma} (\eta^{\alpha\beta}\eta^{\alpha^{'} \gamma^{'}} +
\eta^{\alpha^{'}\beta} \eta^{\alpha \gamma^{'}} ) \nn\\
&+(p-k)^{\gamma^{'}} (\eta^{\alpha\beta}\eta^{\alpha^{'} \gamma } +
\eta^{\alpha^{'}\beta} \eta^{\alpha \gamma} ) .
\eeqa
Substituting this result into the expression (\ref{vecror2tensorvertex})
with the terms in the reversed cyclic orientation
$a,(\alpha,\alpha^{'}),k \leftrightarrow c,(\gamma,\gamma^{'}),q$,
we get:
\beqa
\CV^{\alpha \alpha^{'} \beta \gamma\gamma^{'}}_{a  b c } (k  ,p ,q) =
 ~tr([\lambda^{a } , \lambda^{b}]\lambda^{c})
F^{\alpha \alpha^{'} \beta \gamma\gamma^{'}} (k  ,p ,q)~.
\eeqa
The remaining part of the vertex VTT (\ref{linearcubic}), which is cubic in momenta,
$^{(3)}F$ is
\beqa\label{cubic}
 ^{(3)}F^{\alpha \alpha^{'} \beta \gamma\gamma^{'}} (k  ,p ,q)=
  8 \alpha^{'} [~
-\eta^{\alpha^{'} \gamma^{'}} k^{\alpha}_3 k^{\gamma}_1 k^{\beta}_2
+\eta^{\beta \gamma^{'}} k^{\alpha}_3 k^{\alpha^{'}}_2 k^{\gamma}_1
- \eta^{\beta\alpha^{'}} k^{\alpha }_2   k^{\gamma}_3   k^{\gamma^{'}}_1~]
\eeqa
and can be written in the  form
\beqa\label{cubicver}
&^{(3)}M_{string}= \varepsilon_{\alpha\acute{\alpha}}(k_1) e_{\beta}(k_3)
\varepsilon_{\gamma\gamma^{'}}(k_2) ~ ^{(3)} F^{\alpha \alpha^{'} \beta \gamma\gamma^{'}} (k  ,p ,q)= \\
&8\alpha^{'} [~  {1 \over 2} ~(p \cdot \varepsilon_k  \cdot \varepsilon_q \cdot p)~((q-k) \cdot e_p)
 - (p  \cdot \varepsilon_k \cdot p) (k \cdot \varepsilon_q \cdot e_p)
 +    (q  \cdot \varepsilon_k \cdot e_p) (p \cdot \varepsilon_q \cdot p) ~].\nn
\eeqa
We can calculate its value in the limit considered in section 4, that gives
\beqa
&^{(3)}M(+2,+1,-2)_{string}=\varepsilon_{\alpha\acute{\alpha}}(k_1) e_{\beta}(k_3)
\varepsilon_{\gamma\gamma^{'}}(k_2) ~ ^{(3)}
F^{\alpha \alpha^{'} \beta \gamma\gamma^{'}} (k  ,p ,q)=0,
\eeqa
because $k  \cdot e_q =0,~ p \cdot e_q    =0$.
Its gauge variation   under
(\ref{variationa}) $\delta e_p \sim p$ vanishes
\beqa
&\delta~ ^{(3)}M_{string} = 0 \nn
\eeqa
and under (\ref{variationb}) $\delta \varepsilon_k \sim k \otimes \xi + \xi \otimes k$
vanishes as well
\beqa
&\delta~ ^{(3)}M_{string} =0 ,\nn
\eeqa
because of the same relations $k  \cdot e_q =0,~ p \cdot e_q    =0$.

The cubic in momenta part of the VTT vertex can be associated with
an effective action which have higher derivative terms
constructed by generalized field strength tensors (\ref{tensors}).
The gauge invariant effective action which is cubic in field strength tensors is
\beqa
&\CL_{\partial} =G_{\mu\nu,\lambda} G_{\nu \rho } G_{\rho\mu,\lambda}+
{1 \over 2} G_{\mu\nu} G_{\nu \rho,\lambda\lambda } G_{\rho\mu}.
\eeqa
Indeed, as one can easily check, its gauge variation vanishes
\beqa
&\delta \CL_{\partial} = ([G_{\mu\nu,\lambda}~\xi] +[G_{\mu\nu}~\xi_\lambda]) G_{\nu \rho } G_{\rho\mu,\lambda}+
G_{\mu\nu,\lambda} [G_{\nu \rho }~\xi] G_{\rho\mu,\lambda}+
G_{\mu\nu,\lambda} G_{\nu \rho } ([G_{\rho\mu,\lambda}~\xi]+ [G_{\rho\mu} ~ \xi_{\lambda}])+
\nn\\
&~{1 \over 2}\{[G_{\mu\nu}~\xi] G_{\nu \rho,\lambda\lambda } G_{\rho\mu}+
G_{\mu\nu} ([G_{\nu \rho,\lambda\lambda}~\xi] + 2 [G_{\nu \rho,\lambda}~ \xi_{\lambda}]
+[G_{\nu \rho}~ \xi_{\lambda\lambda}] )  G_{\rho\mu}+
G_{\mu\nu} G_{\nu \rho,\lambda\lambda }
[G_{\rho\mu}~ \xi] \}=0\nn
\eeqa
On the other hand, the second invariant we have found has the form
\beqa
 \CL^{'}_{\partial} =&- Tr(G_{\mu\nu,\lambda} G_{\nu \rho } G_{\rho \mu,\lambda})
+ Tr(G_{\mu\lambda,\lambda} G_{\mu\nu } G_{\nu \rho,\rho})
+Tr(G_{\mu\nu,\lambda} G_{\mu \rho } G_{\rho \lambda,\nu})~+\nn\\
&+Tr(G_{\mu\nu} G_{\mu \rho,\nu}  G_{\rho \lambda,\lambda})+
Tr(G_{\mu\nu} G_{\mu \rho,\lambda}  G_{\rho \lambda,\nu})+
Tr(G_{\mu\nu} G_{\mu \rho,\lambda}  G_{\rho\nu, \lambda})+\nn\\
&+Tr(G_{\mu\lambda,\lambda} G_{\mu \nu,\rho}  G_{\nu\rho })+
Tr(G_{\mu\nu,\lambda} G_{\mu \rho,\lambda}  G_{\rho\nu })+
Tr(G_{\mu\nu,\lambda} G_{\mu \rho,\nu}  G_{\rho \lambda})+\nn\\
&+2Tr(G_{\mu\nu } G_{\nu\lambda, \rho \mu} G_{\rho\lambda})-
Tr(G_{\mu\nu} G_{\nu\rho, \lambda\lambda} G_{\rho \mu})~ .
\eeqa

\section*{\it Acknowledgements}

The work of (I.A.) was supported in part by the European Commission under the ERC Advanced Grant 226371
and in part by the CNRS grant GRC APIC PICS 3747.
The work of (G.S.) was partially supported by ENRAGE (European Network on Random
Geometry), a Marie Curie Research Training Network, contract MRTN-CT-2004-
005616.

\section*{{\it Appendix}}
Let us consider four terms of the interaction vertex VTT. We have
\be
\prod_{i<j}\vert y_i -y_j \vert^{2\alpha^{'}k_ik_j}~ = ~  \vert y_{12} \vert^{2 }~
\ee
because the mass-shell conditions are
$
\alpha^{'} k^{2}_{1}=\alpha^{'} k^{2}_{2}= -1,~   \alpha^{'} k^{2}_{3}=0
$,
implying that
$
k_1 \cdot k_3 = k_1 \cdot k_3 = 0,~~~2 \alpha^{'} k_1 \cdot k_2 =
- \alpha^{'} (k^2_1+k^2_2) = 2.
$
The first term gives
\beqa
&y^2_{12} <:c e^{-\phi} \psi^{\alpha} \dot{X}^{\alpha^{'}}  e^{ik_1  X}(y_1):~
:c e^{-\phi} \psi^{\gamma} \dot{X}^{\gamma^{'}}  e^{ik_2 X}(y_2):~
:c (\dot{X}^{\beta}
-2i \alpha^{'} k_3\cdot \psi \psi^{\beta} )e^{ik_3 X}(y_3) :>=\nn\\
=& y^2_{12}~ y_{12}~ y_{23}~ y_{13}~y^{-1}_{12}~\{~ {\eta^{\alpha \gamma}\over y_{12} } [F^{\alpha^{'} }_{y_1}  F^{\gamma^{'} } _{y_2} F^{\beta}_{y_3}
-2 \alpha^{'}[F^{\alpha^{'} }_{y_1}~ {\eta^{\gamma^{'} \beta}\over y^{2}_{23}}
+F^{\gamma^{'} }_{y_2}~ {\eta^{\alpha^{'} \beta}\over y^{2}_{13}}
+F^{\beta }_{y_3}~ {\eta^{\alpha^{'}\gamma^{'} }\over y^{2}_{12}}]]-\nn\\
&-2 i \alpha^{'}[F^{\alpha^{'} }_{y_1}  F^{\gamma^{'} } _{y_2} - 2 \alpha^{'}
~ {\eta^{\alpha^{'}\gamma^{'} }\over y^{2}_{12}}]
[~ {\eta^{\alpha \beta } k^{\gamma}_{3}\over y_{13}~ y_{23}}-
{\eta^{\gamma \beta } k^{\alpha}_{3}\over y_{13} ~y_{23} }]~ \}=\nn\\
&=(-2 i\alpha^{'})^3 [\eta^{\alpha \gamma } k^{\beta}_{1}
+\eta^{\alpha \beta } k^{\gamma}_{3} +\eta^{ \gamma\beta } k^{\alpha}_{2}]
k^{\alpha^{'}}_{2} k^{\gamma^{'}}_{3} +\nn\\
&+i(2 \alpha^{'})^2 [\eta^{\alpha \gamma } (\eta^{\beta\gamma^{'} } k^{\alpha^{'}}_{2}
+ \eta^{\alpha^{'}\beta } k^{\gamma^{'}}_{3} +\eta^{\alpha^{'}\gamma^{'} } k^{\beta}_{1})
+ \eta^{\alpha^{'}\gamma^{'} } (\eta^{\alpha \beta } k^{\gamma }_{3}
+ \eta^{\beta \gamma} k^{\alpha}_{2} )]
\eeqa
The second term gives
\beqa
& y^2_{12}<c e^{-\phi} \psi^{\alpha} \dot{X}^{\alpha^{'}}  e^{ik_1  X}(y_1)~
c e^{-\phi} \psi^{\gamma} (-2i \alpha^{'}) k_2 \cdot \psi \psi^{\gamma^{'}}
e^{ik_2 X}(y_2)~ c (\dot{X}^{\beta}
-2i \alpha^{'} k_3\cdot \psi \psi^{\beta} )e^{ik_3 X}(y_3) >=\nn\\
=& y^2_{12}~ y_{12}~ y_{23}~ y_{13}~y^{-1}_{12}~(-2 i \alpha^{'})^2  F^{\alpha^{'} }_{y_1}
[{ \eta^{\alpha \gamma} \over y_{12} } (-{k_2\cdot k_3 \eta^{\beta \gamma^{'} }
\over  y^{2}_{23}} +{k^{\beta}_2 k^{\gamma^{'}}_3 \over y^{2}_{23}} )
-  {k^{\alpha}_2 \over y_{12} } (-{k^{ \gamma}_3  \eta^{\beta \gamma^{'} }
\over  y^{2}_{23}} +{ k^{\gamma^{'}}_3 \eta^{\beta \gamma  } \over y^{2}_{23}} )+\nn\\
&+{ \eta^{\alpha \gamma^{'}} \over y_{12} } (-{  k^{\gamma}_3 k^{\beta}_2
\over  y^{2}_{23}} +{k_2\cdot k_3 \eta^{\beta \gamma} \over y^{2}_{23}} )]=\\
&=(-2 i \alpha^{'})^3 k^{\alpha^{'}}_2 [\eta^{\alpha \gamma} k^{\beta}_2 k^{\gamma^{'}}_3
+\eta^{\beta \gamma^{'}} k^{\alpha}_2 k^{\gamma}_3
- \eta^{\beta\gamma} k^{\alpha }_2 k^{\gamma^{'}}_3
-\eta^{\alpha \gamma^{'}} k^{\beta}_2 k^{\gamma}_3
+ k_2\cdot k_3(\eta^{\alpha \gamma^{'}}\eta^{\beta\gamma}
- \eta^{\alpha \gamma}\eta^{\beta \gamma^{'}}  )]\nn
\eeqa
The third  one gives
\beqa
&y^2_{12} <c e^{-\phi} \psi^{\alpha} (-2i \alpha^{'}) k_1 \cdot \psi \psi^{\alpha^{'}} e^{ik_1  X}(y_1)~
c e^{-\phi} \psi^{\gamma} \dot{X}^{\gamma^{'}}  e^{ik_2 X}(y_2)~
c (\dot{X}^{\beta}-2i \alpha^{'} k_3\cdot \psi \psi^{\beta} )e^{ik_3 X}(y_3) >=\nn\\
=& y^2_{12}~ y_{12}~ y_{23}~ y_{13}~y^{-1}_{12}~(-2 i \alpha^{'})^2  F^{\gamma^{'} }_{y_2}
[{ \eta^{\alpha \gamma} \over y_{12} } (-{k_1\cdot k_3 \eta^{\beta \alpha^{'} }
\over  y^{2}_{13}} +{k^{\beta}_1 k^{\alpha^{'}}_3 \over y^{2}_{13}} )
-  {k^{ \gamma}_1 \over y_{12} } (-{k^{\alpha}_3  \eta^{\alpha^{'} \beta }
\over  y^{2}_{13}} +{ k^{\alpha^{'}}_3 \eta^{\alpha \beta  } \over y^{2}_{13}} )+\nn\\
&+{ \eta^{\alpha^{'} \gamma} \over y_{12} } (-{  k^{\beta}_1 k^{\alpha}_3
\over  y^{2}_{13}} +{k_1\cdot k_3 \eta^{\alpha\beta } \over y^{2}_{13}} )]=\\
&=(-2 i \alpha^{'})^3 k^{\gamma^{'}}_3 [\eta^{\alpha \gamma} k^{\beta}_1 k^{\alpha^{'}}_3
-\eta^{\alpha\beta } k^{\alpha^{'}}_3 k^{\gamma}_1
+ \eta^{\beta\alpha^{'}} k^{\alpha }_3 k^{\gamma}_1
-\eta^{\alpha^{'} \gamma} k^{\beta}_1 k^{\alpha}_3
+ k_1\cdot k_3(\eta^{\alpha \beta}\eta^{\alpha^{'}\gamma}
- \eta^{\alpha \gamma}\eta^{\alpha^{'}\beta }  )]\nn
\eeqa
and the last forth term gives
\beqa
& y^2_{12}<c e^{-\phi} \psi^{\alpha} (-2i \alpha^{'}) k_1 \cdot \psi \psi^{\alpha^{'}} e^{ik_1  X}(y_1)~
c e^{-\phi} \psi^{\gamma} (-2i \alpha^{'}) k_2 \cdot \psi \psi^{\gamma^{'}}~
c (\dot{X}^{\beta}-2i \alpha^{'} k_3\cdot \psi \psi^{\beta} )e^{ik_3 X}(y_3) >=\nn\\
& =y^2_{12}~ y_{12}~ y_{23}~ y_{13}~y^{-1}_{12}~(-2 i \alpha^{'})^3
\nn \\
&\{  {F^{\beta }_{y_3} \over (-2 i \alpha^{'})}
 [{ \eta^{\alpha \gamma} \over y_{12} } (-{k_1\cdot k_2 \eta^{\alpha^{'}\gamma^{'} }
\over  y^{2}_{12}} +{k^{\gamma^{'}}_1 k^{\alpha^{'}}_2 \over y^{2}_{12}} )
-{k^{\alpha}_2 \over y_{12} } (-{k^{\gamma}_1  \eta^{\alpha^{'} \gamma^{'} }
\over  y^{2}_{12}} +{ k^{\gamma^{'}}_1 \eta^{\alpha^{'} \gamma  } \over y^{2}_{12}} )
 +{ \eta^{\alpha \gamma^{'}} \over y_{12} } (-{  k^{\gamma}_1 k^{\alpha^{'}}_2
\over  y^{2}_{12}} +{k_1\cdot k_2 \eta^{\alpha^{'}\gamma } \over y^{2}_{12}} )]
 \nn\\
&-{k^{\alpha}_3 \over y_{13}} [{ \eta^{\beta \gamma} \over y_{23} }
(-{k_1\cdot k_2 \eta^{\alpha^{'}\gamma^{'} }
\over  y^{2}_{12}} +{k^{\gamma^{'}}_1 k^{\alpha^{'}}_2 \over y^{2}_{12}} )
-  {k^{\beta}_2 \over y_{23} } (-{k^{\gamma}_1  \eta^{\alpha^{'} \gamma^{'} }
\over  y^{2}_{12}} +{ k^{\gamma^{'}}_1 \eta^{\alpha^{'} \gamma  } \over y^{2}_{12}} )+
 { \eta^{\beta \gamma^{'}} \over y_{23} } (-{  k^{\gamma}_1 k^{\alpha^{'}}_2
\over  y^{2}_{12}} +{k_1\cdot k_2 \eta^{\alpha^{'}\gamma } \over y^{2}_{12}} )]
\nn\\
&-{k^{\alpha^{'}}_3 \over y_{13}} [{ \eta^{\beta \gamma} \over y_{23} }
({k_1\cdot k_2 \eta^{\alpha\gamma^{'} }\over  y^{2}_{12}}
-{k^{\gamma^{'}}_1 k^{\alpha}_2 \over y^{2}_{12}} )
-  {k^{\beta}_2 \over y_{23} } (-{k^{\gamma^{'}}_1  \eta^{\alpha \gamma }
\over  y^{2}_{12}} +{ k^{\gamma}_1 \eta^{\alpha \gamma^{'}  } \over y^{2}_{12}} )+
 { \eta^{\beta \gamma^{'}} \over y_{23} } ({  k^{\gamma}_1 k^{\alpha}_2
\over  y^{2}_{12}} -{k_1\cdot k_2 \eta^{\alpha \gamma } \over y^{2}_{12}} )]
\nn\\
&+{k^{\gamma}_3 \over y_{23}} [{ \eta^{\alpha\beta } \over y_{13} }
(-{k_1\cdot k_2 \eta^{\alpha^{'}\gamma^{'} }
\over  y^{2}_{12}} +{k^{\gamma^{'}}_1 k^{\alpha^{'}}_2 \over y^{2}_{12}} )
-  {k^{\beta}_1 \over y_{13} } (-{k^{\alpha}_2  \eta^{\alpha^{'} \gamma^{'} }
\over  y^{2}_{12}} +{ k^{\alpha^{'}}_2 \eta^{\alpha \gamma^{'}  } \over y^{2}_{12}} )+
 { \eta^{\beta \alpha^{'}} \over y_{13} } (-{  k^{\gamma^{'}}_1 k^{\alpha}_2
\over  y^{2}_{12}} +{k_1\cdot k_2 \eta^{\alpha\gamma^{'} } \over y^{2}_{12}} )]
\nn\\
&+{k^{\gamma^{'}}_3 \over y_{23}} [{ \eta^{\alpha\beta } \over y_{13} }
({k_1\cdot k_2 \eta^{\alpha^{'}\gamma }
\over  y^{2}_{12}} -{k^{\gamma}_1 k^{\alpha^{'}}_2 \over y^{2}_{12}} )
-  {k^{\beta}_1 \over y_{13} } ({k^{\alpha}_2  \eta^{\alpha^{'} \gamma}
\over  y^{2}_{12}} -{ k^{\alpha^{'}}_2 \eta^{\alpha \gamma  } \over y^{2}_{12}} )+
 { \eta^{\beta \alpha^{'}} \over y_{13} } ({  k^{\gamma}_1 k^{\alpha}_2
\over  y^{2}_{12}} -{k_1\cdot k_2 \eta^{\alpha\gamma } \over y^{2}_{12}} )]
\nn\\
&+{k_1 \cdot k_3 \over y_{13}} [{ \eta^{\beta \gamma} \over y_{23} }
({ k^{\alpha^{'}}_2  \eta^{\alpha\gamma^{'}}_1 \over y^{2}_{12}}
-{ k^{\alpha}_2  \eta^{\alpha^{'}\gamma^{'}} \over y^{2}_{12}} )
-  {k^{\beta}_2 \over y_{23} } (-{  \eta^{\alpha \gamma } \eta^{\alpha^{'}\gamma^{'}}
\over  y^{2}_{12}} +{ \eta^{\alpha^{'}\gamma} \eta^{\alpha \gamma^{'}  } \over y^{2}_{12}} )+
 { \eta^{\beta \gamma^{'}} \over y_{23} } ({  k^{\alpha}_2 \eta^{\alpha^{'} \gamma }
\over  y^{2}_{12}} -{k^{\alpha^{'}}_2 \eta^{\alpha \gamma } \over y^{2}_{12}} )]
\nn\\
&-{k_2 \cdot k_3 \over y_{23}} [{ \eta^{\alpha\beta } \over y_{13} }
({ k^{\gamma^{'}}_1  \eta^{\alpha^{'}\gamma}_1 \over y^{2}_{12}}
-{ k^{\gamma}_1  \eta^{\alpha^{'}\gamma^{'}} \over y^{2}_{12}} )
-  {k^{\beta}_1 \over y_{13} } (-{  \eta^{\alpha \gamma } \eta^{\alpha^{'}\gamma^{'}}
\over  y^{2}_{12}} +{ \eta^{\alpha^{'}\gamma} \eta^{\alpha \gamma^{'}  } \over y^{2}_{12}} )+
 { \eta^{\beta \alpha^{'}} \over y_{13} } ({  k^{\gamma}_1 \eta^{\alpha \gamma^{'} }
\over  y^{2}_{12}} -{k^{\gamma^{'}}_1 \eta^{\alpha \gamma } \over y^{2}_{12}} )].
\nn
\}
\eeqa
After some simple algebra, one can get the expressions given in the main text.


\begin{thebibliography}{99}
%\cite{Neveu:1971mu}
\bibitem{Neveu:1971mu}
  A.~Neveu and J.~Scherk,
\emph{ Connection between Yang-Mills fields and dual models,}
  Nucl.\ Phys.\  B {\bf 36} (1972) 155.
  %%CITATION = NUPHA,B36,155;%%


%\cite{Yoneya:1974jg}
\bibitem{Yoneya:1974jg}
  T.~Yoneya,
  \emph{Connection of Dual Models to Electrodynamics and Gravidynamics,}
  Prog.\ Theor.\ Phys.\  {\bf 51} (1974) 1907.
  %%CITATION = PTPKA,51,1907;%%

%\cite{Yoneya:2009ju}
\bibitem{Yoneya:2009ju}
  T.~Yoneya,
 \emph{Gravity from strings: personal reminiscence on early developments,}
  arXiv:0901.0079 [hep-th].
  %%CITATION = ARXIV:0901.0079;%%

%\cite{Scherk:1974ca}
\bibitem{Scherk:1974ca}
  J.~Scherk and J.~H.~Schwarz,
 \emph{Dual Models For Nonhadrons,}
  Nucl.\ Phys.\  B {\bf 81} (1974) 118.
  %%CITATION = NUPHA,B81,118;%%


%\cite{Scherk:1974mc}
\bibitem{Scherk:1974mc}
  J.~Scherk and J.~H.~Schwarz,
 \emph{ Dual Models And The Geometry Of Space-Time,}
  Phys.\ Lett.\  B {\bf 52} (1974) 347.
  %%CITATION = PHLTA,B52,347;%%




%\cite{Green:1987sp}
\bibitem{Green:1987sp}
  M.~B.~Green, J.~H.~Schwarz and E.~Witten,
\emph{ Superstring theory. Vol. 1: Introduction,}
%\href{http://www.slac.stanford.edu/spires/find/hep/www?irn=1755021}{SPIRES entry}
{\it  Cambridge, UK: Univ. Pr. ( 1987) 469 P. ( Cambridge Monographs On Mathematical Physics)}

%\cite{Polchinski:1998rq}
\bibitem{Polchinski:1998rq}
  J.~Polchinski,
\emph{String theory. Vol. 1: An introduction to the bosonic string,}
%\href{http://www.slac.stanford.edu/spires/find/hep/www?irn=4634799}{SPIRES entry}
{\it  Cambridge, UK: Univ. Pr. (1998) 402 p}

%\cite{Witten:1985cc}
\bibitem{Witten:1985cc}
  E.~Witten,
  \emph{Noncommutative Geometry And String Field Theory,}
  Nucl.\ Phys.\  B {\bf 268} (1986) 253.
  %%CITATION = NUPHA,B268,253;%%


%\cite{Thorn:1985fa}
\bibitem{Thorn:1985fa}
  C.~B.~Thorn,
   \emph{Comments On Covariant Formulations Of String Theories,}
  Phys.\ Lett.\  {\bf 159B} (1985) 107;
  %[Addendum-ibid.\  {\bf 160B} (1985) 430].
  %%CITATION = PHLTA,159B,107;%%
    \emph{ String Field Theory,} Phys.\ Rep.\  {\bf 174C} (1989) 1


%\cite{Siegel:1985tw}
\bibitem{Siegel:1985tw}
W.~Siegel and B.~Zwiebach,
\emph{Gauge String Fields,}
Nucl.\ Phys.\ B {\bf 263} (1986) 105.
%%CITATION = NUPHA,B263,105;%%




%\cite{Arefeva:1989cp}
\bibitem{Arefeva:1989cp}
  I.~Y.~Arefeva, P.~B.~Medvedev and A.~P.~Zubarev,
  \emph{ New Representation For String Field Solves
  The Consistence Problem For Open Superstring Field,}
  Nucl.\ Phys.\  B {\bf 341} (1990) 464.
  %%CITATION = NUPHA,B341,464;%%



%\cite{Siegel:1988yz}
\bibitem{Siegel:1988yz}
  W.~Siegel,
 \emph{Introduction to string field theory,}
  arXiv:hep-th/0107094.
  %%CITATION = HEP-TH/0107094;%%

%\cite{Taylor:2003gn}
\bibitem{Taylor:2003gn}
  W.~Taylor and B.~Zwiebach,
   \emph{D-branes, tachyons, and string field theory,}
  arXiv:hep-th/0311017 (see formulas (201) and (202) of section 6.5).
  %%CITATION = HEP-TH/0311017;%%

%\cite{Taylor:2006ye}
\bibitem{Taylor:2006ye}
  W.~Taylor,
  \emph{String field theory,}
  arXiv:hep-th/0605202.
  %%CITATION = HEP-TH/0605202;%%


\bibitem{schwinger}J.Schwinger, \emph{ Particles, Sourses, and Fields}
(Addison-Wesley, Reading, MA, 1970)

%\cite{Weinberg:1964cn}
\bibitem{Weinberg:1964cn}
S.~Weinberg,
\emph{Feynman Rules For Any Spin,}
Phys.\ Rev.\  {\bf 133} (1964) B1318.
%%CITATION = PHRVA,133,B1318;%%

%\cite{Weinberg:1964ev}
\bibitem{Weinberg:1964ev}
S.~Weinberg, \emph{Feynman Rules For Any Spin. 2: Massless Particles,}
Phys.\ Rev.\  {\bf 134} (1964) B882.
%%CITATION = PHRVA,134,B882;%%

%\cite{Weinberg:1964ew}
\bibitem{Weinberg:1964ew}
S.~Weinberg, \emph{Photons And Gravitons In S Matrix Theory: Derivation Of Charge Conservation
And Equality Of Gravitational And Inertial Mass,}
Phys.\ Rev.\  {\bf 135} (1964) B1049.
%%CITATION = PHRVA,135,B1049;%%


\bibitem{chang}S.~J.~Chang, \emph{Lagrange Formulation for Systems with Higher Spin},
Phys.Rev. {\bf 161} (1967) 1308

\bibitem{singh}L.~P.~S.~Singh and C.~R.~Hagen, \emph{Lagrangian formulation for
arbitrary spin. I. The boson case},
Phys.\ Rev.\ {\bf D9} (1974) 898

\bibitem{singh1}L.~P.~S.~Singh and C.~R.~Hagen, \emph{Lagrangian formulation for
arbitrary spin. II. The fermion case}
Phys.\ Rev.\ {\bf D9} (1974) 898, 910

\bibitem{fronsdal}C.Fronsdal, \emph{Massless fields with integer spin,}
Phys.Rev. {\bf D18} (1978) 3624

\bibitem{fronsdal1}J.Fang and C.Fronsdal, \emph{
Massless fields with half-integral spin,}
Phys.\ Rev.\ {\bf D18} (1978) 3630

%\cite{Bengtsson:1983pd}
\bibitem{Bengtsson:1983pd}
A.~K.~Bengtsson, I.~Bengtsson and L.~Brink,
\emph{Cubic Interaction Terms For Arbitrary Spin,}
Nucl.\ Phys.\ B {\bf 227} (1983) 31.

%\cite{Bengtsson:1983pg}
\bibitem{Bengtsson:1983pg}
A.~K.~Bengtsson, I.~Bengtsson and L.~Brink,
\emph{Cubic Interaction Terms For Arbitrarily Extended Supermultiplets,}
Nucl.\ Phys.\ B {\bf 227} (1983) 41.


%\cite{Bengtsson:1986kh}
\bibitem{Bengtsson:1986kh}
  A.~K.~H.~Bengtsson, I.~Bengtsson and N.~Linden,
\emph{Interacting Higher Spin Gauge Fields on the Light Front,}
  Class.\ Quant.\ Grav.\  {\bf 4} (1987) 1333.
  %%CITATION = CQGRD,4,1333;%%

%\cite{Berends:1984rq}
\bibitem{Berends:1984rq}
  F.~A.~Berends, G.~J.~H.~Burgers and H.~van Dam,
\emph{On The Theoretical Problems In Constructing Interactions Involving Higher
  Spin Massless Particles,}
  Nucl.\ Phys.\  B {\bf 260} (1985) 295.
  %%CITATION = NUPHA,B260,295;%%




%\cite{Curtright:1987zc}
\bibitem{Curtright:1987zc}
T.~Curtright, \emph{Generalized Gauge Fields,}
Phys.\ Lett.\  {\bf B165} (1985) 304
%[Erratum-ibid.\  {\bf 60} (1988) 1990].
%%CITATION = PRLTA,60,393;%%



%\cite{Bekaert:2005vh}
\bibitem{Bekaert:2005vh}
  X.~Bekaert, S.~Cnockaert, C.~Iazeolla and M.~A.~Vasiliev,
 \emph{Nonlinear higher spin theories in various dimensions,}
  arXiv:hep-th/0503128.
  %%CITATION = HEP-TH/0503128;%%

%\cite{Vasiliev:2005zu}
\bibitem{Vasiliev:2005zu}
  M.~A.~Vasiliev,
 \emph{Actions, charges and off-shell fields in the unfolded dynamics  approach,}
  Int.\ J.\ Geom.\ Meth.\ Mod.\ Phys.\  {\bf 3} (2006) 37
  %[arXiv:hep-th/0504090].
  %%CITATION = 00436,3,37;%%

%\cite{Francia:2006hp}
\bibitem{Francia:2006hp}
  D.~Francia and A.~Sagnotti,
  \emph{Higher-spin geometry and string theory,}
  J.\ Phys.\ Conf.\ Ser.\  {\bf 33} (2006) 57
  [arXiv:hep-th/0601199].
  %%CITATION = 00462,33,57;%%

%\cite{Bouatta:2004kk}
\bibitem{Bouatta:2004kk}
  N.~Bouatta, G.~Compere and A.~Sagnotti,
 \emph{An introduction to free higher-spin fields,}
  arXiv:hep-th/0409068.
  %%CITATION = HEP-TH/0409068;%%



%\cite{Sagnotti:2005ns}
\bibitem{Sagnotti:2005ns}
  A.~Sagnotti, E.~Sezgin and P.~Sundell,
\emph{On higher spins with a strong Sp(2,R) condition,}
  arXiv:hep-th/0501156.
  %%CITATION = HEP-TH 0501156;%%

\bibitem{Bekaert:2006us}
  X.~Bekaert, N.~Boulanger, S.~Cnockaert and S.~Leclercq,
\emph{On Killing tensors and cubic vertices in higher-spin gauge theories,}
  Fortsch.\ Phys.\  {\bf 54} (2006) 282
  [arXiv:hep-th/0602092].

%\cite{Bekaert:2005jf}
\bibitem{Bekaert:2005jf}
  X.~Bekaert, N.~Boulanger and S.~Cnockaert,
\emph{Spin three gauge theory revisited,}
  JHEP {\bf 0601} (2006) 052
  [arXiv:hep-th/0508048].


%\cite{Metsaev:2005ar}
\bibitem{Metsaev:2005ar}
  R.~R.~Metsaev,
\emph{Cubic interaction vertices of massive and massless higher spin fields,}
  Nucl.\ Phys.\  B {\bf 759} (2006) 147
  %[arXiv:hep-th/0512342].
  %%CITATION = NUPHA,B759,147;%%




%\cite{Paton:1969je}
\bibitem{Paton:1969je}
  J.~E.~Paton and H.~M.~Chan,
\emph{Generalized veneziano model with isospin,}
  Nucl.\ Phys.\  B {\bf 10} (1969) 516.
  %%CITATION = NUPHA,B10,516;%%





%\cite{Kawai:1985xq}
\bibitem{Kawai:1985xq}
  H.~Kawai, D.~C.~Lewellen and S.~H.~H.~Tye,
\emph{A Relation Between Tree Amplitudes Of Closed And Open Strings,}
  Nucl.\ Phys.\  B {\bf 269} (1986) 1.
  %%CITATION = NUPHA,B269,1;%%

%\cite{Antoniadis:2007ta}
\bibitem{Antoniadis:2007ta}
  I.~Antoniadis and S.~Hohenegger,
 \emph{Topological amplitudes and physical couplings in string theory,}
  Nucl.\ Phys.\ Proc.\ Suppl.\  {\bf 171} (2007) 176
  [arXiv:hep-th/0701290].
  %%CITATION = NUPHZ,171,176;%%


%\cite{Antoniadis:2009nv}
\bibitem{Antoniadis:2009nv}
  I.~Antoniadis, S.~Hohenegger, K.~S.~Narain and E.~Sokatchev,
 \emph{ A New Class of N=2 Topological Amplitudes,}
  arXiv:0905.3629 [hep-th].
  %%CITATION = ARXIV:0905.3629;%%

%\cite{Antoniadis:2006mr}
\bibitem{Antoniadis:2006mr}
  I.~Antoniadis, S.~Hohenegger and K.~S.~Narain,
 \emph{ N = 4 topological amplitudes and string effective action,}
  Nucl.\ Phys.\  B {\bf 771} (2007) 40
  [arXiv:hep-th/0610258].
  %%CITATION = NUPHA,B771,40;%%





%\cite{Savvidy:2005fi}
\bibitem{Savvidy:2005fi}
G.~Savvidy,
\emph{Non-Abelian tensor gauge fields: Generalization of Yang-Mills theory},
Phys.\ Lett.\ B {\bf 625} (2005) 341


\bibitem{Savvidy:2005zm}
%\cite{Savvidy:2005zm}
  G.~Savvidy,
 \emph{Non-abelian tensor gauge fields. I,}
  Int.\ J.\ Mod.\ Phys.\ A {\bf 21} (2006) 4931;
  %{\tt[hep-th/0505033]}.
  %%CITATION = IMPAE,A21,4931;


\bibitem{Savvidy:2005ki}
  G.~Savvidy,
  \emph{Non-abelian tensor gauge fields. II,}
  Int.\ J.\ Mod.\ Phys.\ A {\bf 21} (2006) 4959;
  %{\tt [hep-th/0510258]}
  %%CITATION = IMPAE,A21,4959;%%

%\cite{Savvidy:2005vm}
\bibitem{Savvidy:2005vm}
  G.~Savvidy,
 \emph{Non-Abelian tensor gauge fields and higher-spin extension of standard
  model,}
  Fortsch.\ Phys.\  {\bf 54} (2006) 472
  %[arXiv:hep-th/0512012].
  %%CITATION = FPYKA,54,472;%%

%\cite{Savvidy:2008ks}
\bibitem{Savvidy:2008ks}
  G.~Savvidy,
  \emph{Connection between non-Abelian tensor gauge fields and open strings,}
  J.\ Phys.\ A  {\bf 42} (2009) 065403
  %[arXiv:0808.2244 [hep-th]].
  %%CITATION = JPAGB,A42,065403;%%


%\cite{Berends:1981rb}
\bibitem{Berends:1981rb}
  F.~A.~Berends, R.~Kleiss, P.~De Causmaecker, R.~Gastmans and T.~T.~Wu,
  \emph{Single Bremsstrahlung Processes In Gauge Theories,}
  Phys.\ Lett.\  B {\bf 103} (1981) 124.
  %%CITATION = PHLTA,B103,124;%%


%\cite{Dixon:1996wi}
\bibitem{Dixon:1996wi}
  L.~J.~Dixon,
\emph{Calculating scattering amplitudes efficiently,}
  arXiv:hep-ph/9601359.
  %%CITATION = HEP-PH/9601359;%%

%\cite{Parke:1986gb}
\bibitem{Parke:1986gb}
  S.~J.~Parke and T.~R.~Taylor,
 \emph{An Amplitude for $n$ Gluon Scattering,}
  Phys.\ Rev.\ Lett.\  {\bf 56} (1986) 2459.
  %%CITATION = PRLTA,56,2459;%%


%\cite{Witten:2003nn}
\bibitem{Witten:2003nn}
  E.~Witten,
 \emph{Perturbative gauge theory as a string theory in twistor space,}
  Commun.\ Math.\ Phys.\  {\bf 252} (2004) 189
  %[arXiv:hep-th/0312171].
  %%CITATION = CMPHA,252,189;%%

%\cite{Britto:2004ap}
\bibitem{Britto:2004ap}
  R.~Britto, F.~Cachazo and B.~Feng,
 \emph{New recursion relations for tree amplitudes of gluons,}
  Nucl.\ Phys.\  B {\bf 715} (2005) 499
  %[arXiv:hep-th/0412308].
  %%CITATION = NUPHA,B715,499;%%

%\cite{Britto:2005fq}
\bibitem{Britto:2005fq}
  R.~Britto, F.~Cachazo, B.~Feng and E.~Witten,
 \emph{Direct proof of tree-level recursion relation in Yang-Mills theory,}
  Phys.\ Rev.\ Lett.\  {\bf 94} (2005) 181602
  [arXiv:hep-th/0501052].
  %%CITATION = PRLTA,94,181602;%%



%\cite{Benincasa:2007xk}
\bibitem{Benincasa:2007xk}
  P.~Benincasa and F.~Cachazo,
\emph{Consistency Conditions on the S-Matrix of Massless Particles,}
  arXiv:0705.4305 [hep-th].









\end{thebibliography}
\end{document}